\ifx\targetformat\undefined	% seperate main text / supplemental
\def\clsstyle{prl} 

\newcommand{\appref}[1]{the Supplemental Materials}
\newcommand{\arxivtext}[1]{}
\newcommand{\prltext}[1]{#1}
\else		% for arXiv
\def\clsstyle{pra} 

\newcommand{\appref}[1]{App.~\ref{#1}}
\newcommand{\arxivtext}[1]{#1}
\newcommand{\prltext}[1]{}

\fi

\documentclass[aps,\clsstyle,reprint,twocolumn,showpacs,superscriptaddress]{revtex4-1}

\usepackage{amsmath, amssymb, bm, braket, dsfont, times, amsthm}
\usepackage{graphicx}
\usepackage[breaklinks,colorlinks,allcolors=blue]{hyperref}
\usepackage[normalem]{ulem} %for \sout

\newtheorem{theorem}{Theorem}
\newtheorem{lemma}{Lemma}
\newtheorem{corollary}{Corollary}

\usepackage[usenames,dvipsnames]{color}			% Only used for author comments
\definecolor{Zcolour}{rgb}{0.992, 0.588, 0.22}
\definecolor{dkgreen}{rgb}{0,0.5,0}
\definecolor{purple}{rgb}{0.5,0,0.5}

\usepackage[caption=false]{subfig}

\usepackage{tikz}
\definecolor{newblue}{RGB}{112,178,255}
\definecolor{neworange}{RGB}{255,204,112}
\usetikzlibrary{arrows,calc}
\tikzset{
	>=stealth',
	help lines/.style={dashed, thick},
	important line/.style={thick},
	connection/.style={thick, dotted},
}

\begin{document}

\title{Topology and edge modes in quantum critical chains}
\author{Ruben Verresen}
\affiliation{Department of Physics, T42, Technische Universit\"at M\"unchen, 85748 Garching, Germany}
\affiliation{Max-Planck-Institute for the Physics of Complex Systems, 01187 Dresden, Germany}
\author{Nick G. Jones}
\affiliation{School of Mathematics, University of Bristol, BS8 1TW Bristol, United Kingdom}
\author{Frank Pollmann}
\affiliation{Department of Physics, T42, Technische Universit\"at M\"unchen, 85748 Garching, Germany}
%\affiliation{Department of Physics, T42, Technische Universit\"at M\"unchen, James-Franck-Str. 1, 85748 Garching, Germany}
\date{\today}                                       

\begin{abstract}
	We show that topology can protect exponentially localized, zero energy edge modes at critical points between one-dimensional symmetry protected topological phases. This is possible even without gapped degrees of freedom in the bulk ---in contrast to recent work on edge modes in gapless chains. We present an intuitive picture for the existence of these edge modes in the case of non-interacting spinless fermions with time reversal symmetry (BDI class of the tenfold way). The stability of this phenomenon relies on a topological invariant defined in terms of a complex function, counting its zeros and poles inside the unit circle. This invariant can prevent two models described by the \emph{same} conformal field theory (CFT) from being smoothly connected. A full classification of critical phases in the non-interacting BDI class is obtained: each phase is labeled by the central charge of the CFT, $c \in \frac{1}{2}\mathbb N$, and the topological invariant, $\omega \in \mathbb Z$. Moreover, $c$ is determined by the difference in the number of edge modes between the phases neighboring the transition. Numerical simulations show that the topological edge modes of critical chains can be stable in the presence of interactions and disorder.
\end{abstract}
\maketitle

\textbf{Introduction.} Topology is fundamental to characterizing quantum phases of matter in the absence of local order parameters \cite{Wen16}. In one spatial dimension, such zero-temperature phases have topological invariants generically protecting \emph{edge modes}, i.e. zero energy excitations localized near boundaries. These phases are usually referred to as topological insulators or superconductors when non-interacting \cite{Hasan10} and as symmetry protected topological phases when interacting \cite{Gu09,Pollmann12}. The topological invariants of one-dimensional systems require the presence of a symmetry and have been classified for phases with a gap above the ground state \cite{Schnyder08,Kitaev09,Ryu10,Fidkowski11class,Turner11class,Chen10,Schuch11}.

Conventional wisdom says topological edge modes \emph{require} a bulk gap. Recently, there has been work on gapless phases hosting edge modes \cite{Kestner11,Cheng11,Fidkowski11longrange,Sau11,Kraus13,Keselman15,Iemini15,Lang15,Montorsi17,Ruhman17,Scaffidi17_preprint,Jiang17_preprint,Zhang17}, but when their localization is exponential, it is attributed to gapped degrees of freedom (meaning there are exponentially decaying correlation functions).

We indicate that this picture is at odds with the critical points between topological superconductors in the BDI class (noninteracting spinless fermions with time-reversal symmetry) \cite{Altland97}.
Curiously, a 2001 work by Motrunich, Damle and Huse with different aims implied that some of these transitions host topological edge modes \cite{Motrunich01}.
This phenomenon and its consequences have not been explored since. It is of particular importance given the recent interest in the interplay between topology and criticality, since ---as we will argue--- the bulk has \emph{no} gapped degrees of freedom.

After reviewing the gapped phases of the BDI class, we present an example of a critical chain with edge modes. Subsequently, we identify a topological invariant in terms of a complex function ---distinct from the transfer matrix approach \cite{Brouwer98,Fulga11} of Ref.~\cite{Motrunich01}--- which we prove counts the edge modes (Theorem \ref{thm}). Similar concepts appear in the literature on \emph{gapped} topological phases \cite{deGottardi13,Das14,Mandal15}. The idea of classifying such critical phases is then explored. We find that in the BDI class, chains with the \emph{same} conformal field theory (CFT) can be smoothly connected \emph{if and only if} the topological invariant coincides (Theorem \ref{thm_class}). Moreover, the CFT is determined by the change in topological invariant upon crossing the transition (Theorem \ref{thm_transition}). Finally, we numerically demonstrate that topological edge modes at criticality can survive disorder and interactions.

\begin{figure}
	\centering
	\includegraphics[scale=.85]{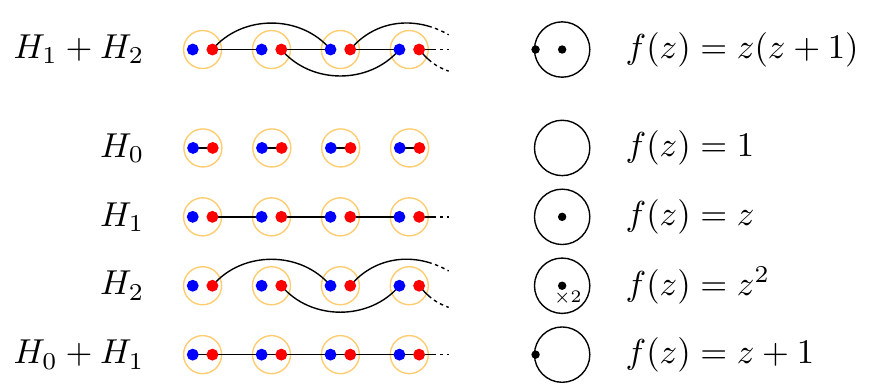}
	\caption{Representation of the critical Hamiltonian $H_1+H_2$ with its edge mode (each fermionic site is decomposed into Majorana modes: $\gamma$ (blue) and $\tilde \gamma$ (red), a bond signifies a term in the Hamiltonian). Also shown are the gapped Hamiltonians $H_\alpha$ ($\alpha = 0$ is trivial and $\alpha=1$ is the Kitaev chain). $H_0 + H_1$ is the standard critical Majorana chain. The associated complex function $f(z)$ (Eq.~\eqref{function}) and zeros in the complex plane are shown. \label{fig:example}}
\end{figure}

\textbf{Example.} We illustrate how a critical phase ---without gapped degrees of freedom--- can have localized edge modes. First, we decompose every fermionic site $c_n,c_n^\dagger$ into two Majorana modes: $\gamma_n = c_n^\dagger + c_n$ and $\tilde \gamma_n = i(c_n^\dagger - c_n)$.
%\begin{equation} \label{def:majorana}
%\gamma_n = c_n^\dagger + c_n \qquad \qquad \tilde \gamma_n = i(c_n^\dagger - c_n)\; .
%\end{equation}
The former is real ($T \gamma_n T = \gamma_n$ where $T$ is complex conjugation in the occupation basis) and the latter imaginary ($T \tilde \gamma_n T = -\tilde \gamma_n$). These Hermitian operators anticommute and square to unity. 

We define the $\alpha$-chain \cite{Verresen17,deGottardi13,Niu12}:
\begin{equation}
H_\alpha = \frac{i}{2} \sum_n \tilde \gamma_n \gamma_{n+\alpha} \qquad (\alpha \in \mathbb Z) \; .
\end{equation}
These \emph{gapped} chains are illustrated in Fig.~\ref{fig:example}. For $\alpha = 1$, it is the Kitaev chain with Majorana edge mode $\gamma_1$ \cite{Kitaev01}. $H_\alpha$ has $|\alpha|$ Majorana zero modes per edge and can be thought of as a stack of Kitaev chains. The edge modes survive quadratic, $T$-preserving perturbations due to chirality: if \emph{real} modes prefer to couple to their \emph{left}, some remain decoupled at the left edge. This can be quantified by a topological winding number \cite{Asboth15} counting edge modes, meaning each $H_\alpha$ represents a distinct phase of matter. The $2|\alpha|$ zero modes imply a $2^{|\alpha|}$-fold degeneracy, with finite-size gap $\sim e^{-L/\xi}$ when the modes are exponentially localized.

Consider now a critical point between the phases defined by fixed point Hamiltonians $H_1$ and $H_2$, namely $H_1+ H_2$. Despite it being critical, Fig.~\ref{fig:example} shows a localized Majorana edge mode. Nevertheless, there is no local operator $\mathcal O$ with $\langle \mathcal O_n \mathcal O_m \rangle \sim e^{-|n-m|/\xi}$ \footnote{One can also imagine a box function, corresponding to $\xi = 0$.}. Indeed, for periodic boundary conditions, shifting $\gamma_n \to \gamma_{n-1}$ (which one \emph{cannot} smoothly implement in a local and $T$-preserving way) maps $H_1 + H_2$ to $H_0 + H_1$, the well-studied critical Majorana chain described by a CFT with central charge $c=\frac{1}{2}$ \cite{CFT}.

In the next section, we demonstrate the edge mode's stability: a topological invariant protects it. Similar to the winding number for gapped phases, it quantifies the couplings' chirality. In short: we associate to every chain a complex function $f(z)$ (illustrated in Fig.~\ref{fig:example}) whose number of zeros (minus poles) in the unit disk counts the edge modes.

\textbf{Topology and edge modes in the BDI class.} Consider the full BDI class: chains of non-interacting spinless fermions with time reversal symmetry $T$ (defined above). The aforementioned 
$\{H_\alpha\}_{\alpha \in \mathbb Z}$ form a basis for arbitrary translation invariant Hamiltonians in this class:
\begin{equation}
H_\textrm{BDI} =  \frac{i}{2}\sum_{\alpha=-\infty}^{+\infty} t_\alpha \; \left( \sum_{n \in \textrm{sites}} \tilde \gamma_n \gamma_{n+\alpha} \right) = \sum_{\alpha} t_\alpha H_\alpha \; .\label{ham:BDI}
\end{equation}
We take $t_\alpha$ to be non-zero for only a finite number of $\alpha$ (i.e. $H$ is finite range). Time reversal symmetry forbids terms of the form $i\gamma_n \gamma_m$ and Hermiticity requires $t_\alpha \in \mathbb R$. We mainly work in this translation invariant setting, but the effects of unit cells and disorder are addressed.

$H_\textrm{BDI}$ is determined by the list of numbers $t_\alpha$, or equivalently, by its Fourier transform, $f(k) := \sum_\alpha t_\alpha e^{ik\alpha}$. It is efficiently diagonalized: if $f(k) = \varepsilon_k e^{i\varphi_k}$ (with $\varepsilon_k,\varphi_k \in \mathbb R$), then a Bogoliubov rotation over the angle $\varphi_k$ diagonalizes $H_\textrm{BDI}$, with single-particle spectrum $\varepsilon_k$ 
\ifx\targetformat\undefined
	\cite{suppl}.
\else
	\footnote{See Appendix \ref{app:solution} for details.}.
\fi

\begin{figure}
	\centering
	\includegraphics[scale=.75]{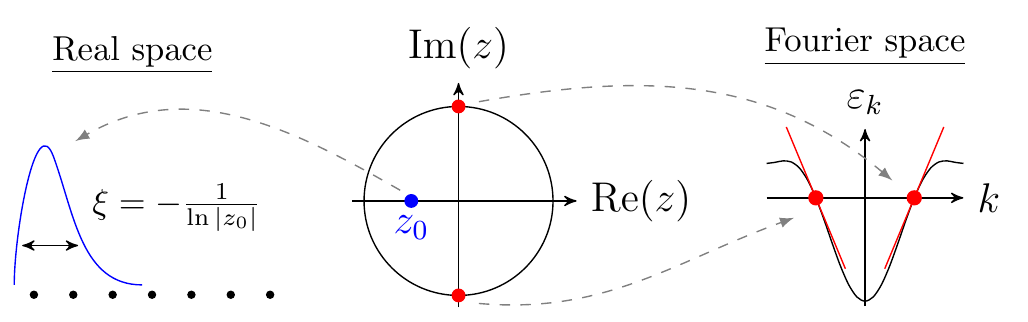}
\caption{The middle figure shows the zeros of $f(z)$. The zero $z_0$ within the disk (blue) corresponds to an edge mode (per edge) with localization length $\xi = -\frac{1}{\ln |z_0|}$. Each zero on the unit circle (red) implies a massless Majorana field in the low-energy limit ($c=\frac{1}{2}$).\label{fig:complex}}
\end{figure}

In this language, the invariant for gapped phases is simply the winding number of $f(k)$ around the origin: since $\varepsilon_k$ is nonzero, the phase $e^{i\varphi_k}$ is a well-defined function from $S^1$ to $S^1$. This fails when the system is gapless, but can be repaired using complex analysis. First, interpret the function $f(k)$ as living on the unit circle in the complex plane ---abusing notation, write $f\left(z=e^{ik}\right)$--- with \emph{unique} analytic continuation
\begin{equation} \label{function}
f(z) = \sum_{\alpha=-\infty}^{\infty} t_\alpha \; z^\alpha \; .
\end{equation}
Now, $f$ is a function $\mathbb C\setminus\{0\} \to \mathbb C$ with a pole at the origin when $t_{\alpha} \neq 0$ for some $\alpha<0$. If it has no zero \emph{on} the unit circle (i.e. the system is gapped), then \emph{Cauchy's argument principle} says that the winding number defined above equals the number of zeros ($N_z$, including degree) minus the order of the pole ($N_p$) within the unit disk. If at least one zero lies on the unit circle, the aforementioned winding number breaks down ---the quantity $N_z - N_p$, however, remains well-defined! Perturbing $H_\textrm{BDI}$ smoothly moves the zeros of $f(z)$ around, and changing the support of $t_\alpha$ produces or destroys zero-pole pairs at the origin or infinity.
Hence, by continuity, $N_z-N_p$ cannot change without affecting the number of zeros on the unit circle. This would change the bulk physics: every (non-degenerate) zero $e^{ik_0}$ of $f(z)$ implies that $\varepsilon_k \sim k-k_0$, contributing a massless Majorana fermion (with central charge $c=\frac{1}{2}$) to the CFT (see Fig.~\ref{fig:complex}).

Hence $\omega := N_z - N_p$ (\emph{strictly within} the unit disk) defines a topological invariant, both for gapped and gapless chains. We now show its physical significance: if $\omega > 0$, it counts the Majorana zero modes which are exponentially localized on the boundary. Moreover, the localization lengths are given by the zeros of $f(z)$. Fig.~\ref{fig:complex} illustrates this, with the precise statement being:

\begin{theorem} \label{thm} If the topological invariant $\omega> 0$, then
	\begin{enumerate}
		\item each boundary has $\omega$ Majorana zero modes,
		\item the modes have localization length $\xi_i = -\frac{1}{\ln|z_i|}$ where $\{ z_i \}$ are the $\omega$ largest zeros of $f(z)$ within the unit disk,
		\item the modes on the left (right) are real (imaginary).
	\end{enumerate} 
	
	If $\omega < -2c$ (where $c =$ half the number of zeros on the unit circle), the left (right) boundary has $|\omega+2c|$ imaginary (real) Majorana modes with localization length $\xi_i = \frac{1}{\ln|z_i|}$, with $\{ z_i \}$ the $|\omega+2c|$ smallest zeros outside the unit disk.
	
	For any other value of $\omega$, no localized edge modes exist.
\end{theorem}
Before we outline the proof, note that in the gapped case $c$ is zero,
with $|\omega|$ correctly counting edge modes. At criticality, $2c$ counts the zeros on the unit circle, and if these are non-degenerate, the bulk is a CFT with central charge $c$. However, if $f(z)$ has a zero $e^{ik_0}$ with multiplicity $m$, then $\varepsilon_{k} \sim (k-k_0)^m$, implying a dynamical critical exponent $z_\textrm{dyn} = m$.

If $\omega > 0$, we construct for each $z_i$ (defined above) a real edge mode $\gamma_\textrm{left}^{(i)} = \sum_{n \geq 1} b_{n}^{(i)} \gamma_n$ by requiring $[\gamma_\textrm{left}^{(i)},H] = 0$. This gives constraints $\sum_{a\geq 1} b_a^{(i)} \; t_{a-n} = 0$ for $n \geq 1$, leading to standard solvable recurrence relations, with the function $f(z)$ appearing as the characteristic polynomial. If $N_p=0$, the solution is simply $b_n^{(i)} \propto z_i^n$ (hence $\big|b_n^{(i)}\big| \sim e^{-n/\xi_i}$), while $N_p > 0$ modifies $b_n^{(i)}$ without affecting its asymptotic form. 
\ifx\targetformat\undefined
Details are treated in the supplemental material.
\else
Details are treated in Appendix \ref{app:proof}.
\fi
The case $\omega < -2c$ follows by noting that inverting left and right effectively implements $f(z) \leftrightarrow f \left(\frac{1}{z}\right)$ and one can show that this changes the topological invariant $\omega \leftrightarrow -\omega-2c$. This completes the proof. Note that the exponential localization implies that the commutator ---and hence the energy gap of the edge mode--- is exponentially small for finite systems.

We can now appreciate the right-hand side of Fig.~\ref{fig:example}, showing for each Hamiltonian the function $f(z)$ and its zeros. The two critical Hamiltonians indeed have a zero on the unit circle, and the edge modes are counted by the zeros strictly within the unit disk. Hence, the edge mode of $H_1 + H_2$ is stable: the zero will stay within the unit disk for small perturbations.

\begin{figure}[h]
	\includegraphics[scale=.9,trim=0cm 0.5cm 0.1cm 0cm]{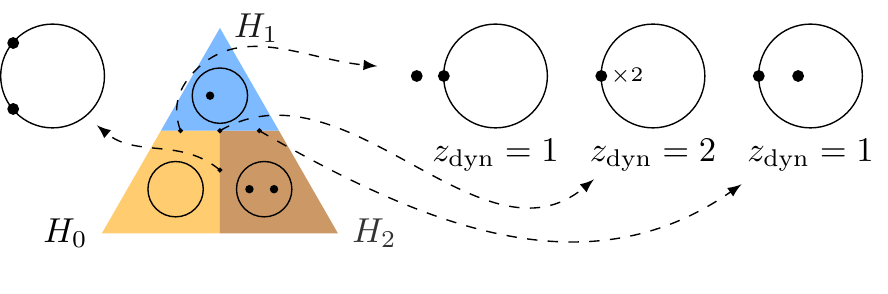}
	\caption{Phase diagram illustrating how critical points with the same CFT description but different topological invariants cannot be connected: interpolating $H_0+H_1$ and $H_1+H_2$ induces a point where the dynamical critical exponent $z_\textrm{dyn}$ changes discontinuously. \label{fig:phases}}
\end{figure}
\textbf{Classifying critical phases.} In this section, we use the above framework to answer two related questions: (1) ``\emph{What is the classification of the critical phases within the BDI class?}'', and (2) ``\emph{Given two gapped phases, what is the universality class of the critical point between them?}''.

We define two Hamiltonians to be in the same phase if and only if they are connected by a path of local Hamiltonians (within the symmetry class) along which the low-energy description of the bulk changes smoothly. (This is different from the notion of Furuya and Oshikawa \cite{Furuya17} where CFTs are in the same phase if a renormalization group flow connects them.) Hamiltonians described by CFTs with different central charges are automatically in distinct phases by the $c$-theorem \cite{Zamolodchikov1986}. If, on the other hand, both Hamiltonians have the same CFT description, one might expect them to be in the same phase. However, we have shown that $H_1+H_2$ and $H_0+H_1$ have different topological invariants (respectively, $\omega= 1$ and $\omega=0$), yet the same CFT description.
Hence, they cannot be connected within the BDI class, as illustrated in Fig.~\ref{fig:phases}.

This is illustrative of the general case. We have seen that any translation invariant Hamiltonian $H_\textrm{BDI}$ can be identified with a complex function $f(z)$, with its set of zeros and poles, as in Fig.~\ref{fig:complex}. Conversely, the zeros and poles uniquely identify $f(z)$ and hence the Hamiltonian! More precisely, let $\{ z_i \}$ be the list of (distinct) zeros of Eq.~\eqref{function} with a corresponding list of multiplicities $\{m_i\}$, and let $N_p$ be the order of the pole at the origin. By the fundamental theorem of algebra, this uniquely identifies the meromorphic function $f(z) = \pm \frac{1}{z^{N_p}} \prod_i (z-z_i)^{m_i}$ up to a positive multiplicative scalar. This correspondence between a BDI Hamiltonian and a picture of zeros and poles reduces the classification to an exercise in geometric insight. 

Let us focus on the physically interesting case where the bulk is described by a CFT ---i.e. the zeros on the unit circle have multiplicity one. There is only one rule restricting the movement of the zeros on the unit circle: since $t_\alpha$ is real, the zeros of $f(z)$ are real or come in complex-conjugate pairs. This means we \emph{cannot} move a zero off the real axis. However, we \emph{can} bring $f(z)$ to a \emph{canonical form} where the zeros are equidistantly distributed on the unit circle with mirror symmetry about the real axis. There are only two such patterns, given by the solutions of $z^{2c} =\pm 1$. Thus, we can always tune to the canonical form $f(z) = \pm \left(z^{2c}\pm 1 \right) z^\omega$. Hence, for a given nonzero $c \in \frac{1}{2} \mathbb N$ and $\omega \in \mathbb Z$, there are $\mathbb Z_2 \times \mathbb Z_2$ translation invariant phases, labeled by the two signs. These signs are protected by translation symmetry: the first encodes the spatially modulating sign of correlations, the second whether $\varepsilon_k = |f\left(e^{ik}\right)|$ vanishes at time-reversal invariant momenta.

Hence, allowing for paths with unit cells, two systems can be smoothly connected \emph{if} they have identical $c$ and $\omega$ \cite{suppl}. To confirm that this is a \emph{necessary} condition, we extend $\omega$ to systems with an $N$-site unit cell, $H = \frac{i}{2} \sum \bm{\tilde \gamma}^T_n T_\alpha \bm{\gamma}_{n+\alpha}$ where $T_\alpha \in \mathbb R^{N\times N}$. Defining $f(z) = \det \left( \sum_\alpha T_\alpha z^\alpha\right)$, then analogous to before, one can show that $|f\left(e^{ik}\right)|$ is the \emph{product} of the energy bands $\varepsilon_k^{n=1,\cdots,N}$ \cite{suppl}. Thus, $\omega = N_z-N_p$ cannot change without a bulk transition.

For gapped phases, $\omega$ is known to be additive under stacking \cite{Asboth15}. Our extension to critical systems still satisfies this property (as does $c$ \cite{CFT}). Moreover, the classification straightforwardly generalizes to stackings, with a small caveat. E.g., a Kitaev chain stacked onto $H_0 + H_1$ has the same invariants as $H_1 + H_2$ ($c = \frac{1}{2}$ and $\omega=1$). One might expect these to be in the same phase, but the latter has no gapped degrees of freedom. This is resolved by adding a decoupled trivial chain ---after which one can indeed connect them. Details are in the supplemental material \cite{suppl}.

We thus obtain a \emph{semigroup} classification (`\emph{semi}' because $c$ cannot decrease under stacking):

\begin{theorem} \label{thm_class} The phases in the BDI class described in the bulk by a CFT and obtained by deforming a translation invariant Hamiltonian (or a stacking thereof) with an arbitrary unit cell, are classified by the semigroup $\mathbb N \times \mathbb Z$: they are labeled by the central charge $c \in \frac{1}{2} \mathbb N$ and the topological invariant $\omega \in \mathbb Z$.
	
Translation invariance gives an extra $\mathbb Z_2$ invariant when $c=0$ and an extra $\mathbb Z_2 \times \mathbb Z_2$ invariant when $c \neq 0$.
\end{theorem}

The second question, regarding the phase transition between two topological phases, is straightforwardly addressed. By continuity, the difference in winding number between two gapped phases is the number of zeros that must cross the unit circle at the transition, i.e.:
\begin{theorem} \label{thm_transition} A phase transition between two gapped phases with winding numbers $\omega_1$ and $\omega_2$ obeys $c \geq \frac{|\omega_1 - \omega_2|}{2}$.
\end{theorem}
As before, $c$ should be understood as counting half the number of zeros on the unit circle. If these zeros are non-degenerate, the bulk is a CFT with central charge $c$. Generically, $c$ \emph{equals} $\frac{|\omega_1 - \omega_2|}{2}$, but one can fine-tune the transition with zeros bouncing off the unit circle \footnote{E.g. $f(z) = (z-b-1)(z+|b|+1)$ at $b=0$ is a CFT with $c=1$ between the trivial phase and the Kitaev chain phase.}. This theorem proves a special case of a recent conjecture concerning all transitions between one-dimensional symmetry protected topological phases \cite{Verresen17}.

\begin{figure}
	\includegraphics[scale=.68]{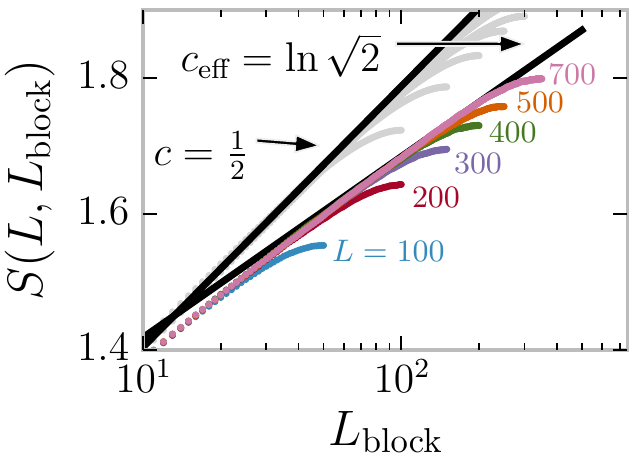}\llap{\parbox[b]{3.3in}{a)\\\rule{0ex}{1.2in}}}
	\hspace{5pt}
	\includegraphics[scale=.68]{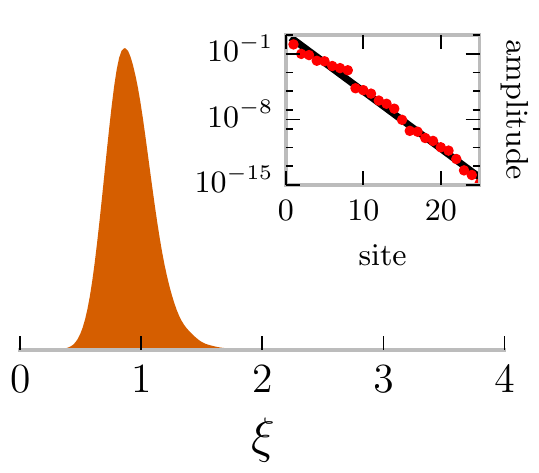}\llap{\parbox[b]{2.8in}{b)\\\rule{0ex}{1.2in}}}
	\caption{Phase transition at strong disorder: (a) entanglement scaling (averaging $10^5$ states) suggests an infinite randomness fixed point with $c_\textrm{eff} = \ln \sqrt{2}$ (black lines guide the eye; gray is the clean case), (b) distribution of edge mode localization length over disorder realizations (inset: edge mode for one realization).\label{fig:disorder}}
\end{figure}

In the remainder, we demonstrate that topological edge modes in critical chains can survive disorder and interactions.

\textbf{Disorder.} We consider $H = \frac{i}{2} \sum_{\alpha=0}^3 \sum_n t_{\alpha}^{(n)} \; \tilde \gamma_n \gamma_{n+\alpha}$. The clean model $t_1^{(n)} = t_2^{(n)} = 1$ and $t_0^{(n)} = t_3^{(n)} = a$ where $-1<a<\frac{1}{3}$, is critical with $c=\frac{1}{2}$ and $\omega = 1$. This reduces to $H_1 + H_2$ when $a=0$. We now introduce strong disorder: $t_1^{(n)},t_2^{(n)}$ $\left(t_0^{(n)},t_3^{(n)}\right)$ are drawn independently from the flat distribution on $[0,1]$ $\left( [-0.5,0]\right)$.

We confirm the system flows to the infinite randomness fixed point with effective central charge $c_\textrm{eff} = \ln \sqrt{2}$ \cite{Motrunich01,Refael04,Laflorencie05}. We diagonalize periodic systems of size $L$, calculating the entanglement entropy $S(L,L_\textrm{block})$ of a region of length $L_\textrm{block}$. The average is predicted to obey the asymptotic scaling $S \sim \frac{c_\textrm{eff}}{3} \ln L_\textrm{block}$ (for $1\ll L_\textrm{block} \ll \frac{L}{2}$), as shown in Fig.~\ref{fig:disorder}(a).

In the presence of open boundary conditions, we observe one Majorana edge mode per boundary. These are exponentially localized, with Fig.~\ref{fig:disorder}(b) showing the distribution of localization lengths over different disorder realizations. The inset gives a generic example, plotting $|b_n|$ where the edge mode is $\gamma_\textrm{left} = \sum_{n=1}^L b_n \gamma_n$.

\textbf{Interactions.} Interactions can have interesting effects, with the gapped classification reducing from $\mathbb Z$ to $\mathbb Z_8$ \cite{Fidkowski10interaction,Fidkowski11class,Turner11class}. Here we simply show that an interacting critical point between the Kitaev chain and $H_2$,
\begin{equation} \label{ham:interaction}
H = H_1 + H_2 + U \sum_{n=1}^L \gamma_{n} \gamma_{n+1} \gamma_{n+2} \gamma_{n+3} + (\gamma \leftrightarrow \tilde \gamma),
\end{equation}
has localized edge modes. The critical point will not shift when $U\neq 0$ since \eqref{ham:interaction} is self-dual under $\gamma_n \to \gamma_{3-n}$ and $\tilde \gamma_n \to \tilde \gamma_{-n}$.

\begin{figure}
	\includegraphics[scale=.62]{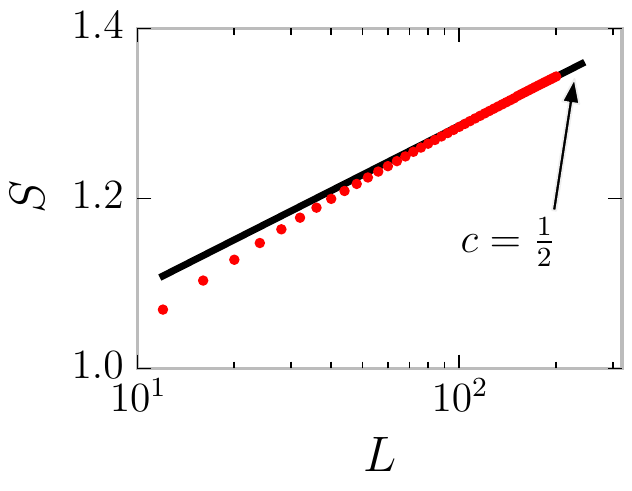}\llap{\parbox[b]{3in}{a)\\\rule{0ex}{1.1in}}}
	\hspace{5pt}
	\includegraphics[scale=.62]{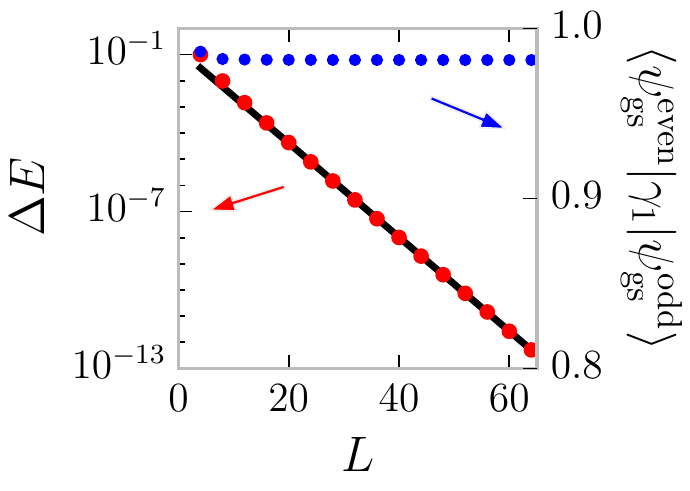}\llap{\parbox[b]{3.2in}{b)\\\rule{0ex}{1.1in}}}
	\caption{Finite-size scaling for the interacting Hamiltonian \eqref{ham:interaction} with open boundaries and $U = 0.3$: (a) the bulk is described by the $c=\frac{1}{2}$ Majorana CFT (black line guides the eye), (b) energy splitting between fermionic parity sectors is exponentially small, $ \Delta E \sim e^{-L/\xi}$ with $\xi \approx 2.42$. The two ground states are related by an edge mode. \label{fig:interaction}}
\end{figure}

We use the density matrix renormalization group method \cite{White92} to perform finite-size scaling with open boundaries for $U=0.3$. (Convergence was reached for system sizes shown with bond dimension $\chi = 60$.) In Fig.~\ref{fig:interaction}(a), we confirm that the system remains critical by using the CFT prediction \cite{Calabrese04} for the the entanglement entropy of a bipartition into two equal halves of length $\frac{L}{2}$, namely $S \sim \frac{c}{6} \ln L$.

Fig.~\ref{fig:interaction}(b) shows the ground state degeneracy with open boundary conditions. These states differ only near the edge since $\langle \psi_\textrm{gs}^\textrm{even} | \gamma_1 | \psi_\textrm{gs}^\textrm{odd} \rangle$ is finite as $L \to \infty$.

\textbf{Conclusion.} We have shown that any two gapped phases in the BDI class with winding numbers $\omega_1>\omega_2>0$ are separated by a critical point with $\omega_2$ topological edge modes $\left(\textrm{and central charge }c=\frac{\omega_1-\omega_2}{2}\right)$. We have characterized such phases within this class in terms of the zeros and poles of an associated complex function.

Unlike gapped phases, these critical chains do not have the usual bulk-boundary correspondence (i.e. edge modes implying degenerate Schmidt values when bipartitioning a periodic system \cite{Pollmann10}), as suggested by Fig.~\ref{fig:example}. Hence, how the topology is reflected in the entanglement is an open question. The study of possible string orders using Toeplitz determinants constitutes a forthcoming work \cite{Jonesfuture}. 

A natural question is how to extend this to the other one-dimensional symmetry classes. We note that our analysis also applies to AIII (identified with a \emph{subset} of BDI, see e.g. Ref.~\onlinecite{Verresen17}) leading to an $\mathbb N\times \mathbb Z$ classification with \emph{integer} central charge. We similarly expect (but do not argue) an $\mathbb N \times \mathbb Z$ classification for CII, and a \emph{single} critical phase \footnote{In principle we should add `per central charge', but only the smallest allowed one will be stable.} for the other non-trivial classes. Furthermore, the classification of interacting critical chains should be found. Similar to the gapped case, a tensor-network approach may prove insightful, perhaps using the multi-scale entanglement renormalization ansatz (MERA) \cite{Vidal07} or infinite-dimensional matrix product states \cite{Cirac10}. The generalization to higher dimensions is an open-ended frontier.

\textbf{Acknowledgements.} The authors would like to thank Erez Berg, Ignacio Cirac, David Huse, Jon Keating, Olexei Motrunich, Tibor Rakovszky, Jonathan Robbins and Guifre Vidal for illuminating discussions. RV and FP thank Roderich Moessner for discussions and collaboration on related topics. RV has been supported by the German Research Foundation (DFG) through the Collaborative Research Centre SFB 1143 and FP acknowledges support from Research Unit FOR 1807 through grant no. PO 1370/2-1 and from the Nanosystems Initiative Munich (NIM) by the German Excellence Inititiative.

\bibliography{paper_v2.bbl}

\begin{thebibliography}{51}%
\makeatletter
\providecommand \@ifxundefined [1]{%
 \@ifx{#1\undefined}
}%
\providecommand \@ifnum [1]{%
 \ifnum #1\expandafter \@firstoftwo
 \else \expandafter \@secondoftwo
 \fi
}%
\providecommand \@ifx [1]{%
 \ifx #1\expandafter \@firstoftwo
 \else \expandafter \@secondoftwo
 \fi
}%
\providecommand \natexlab [1]{#1}%
\providecommand \enquote  [1]{``#1''}%
\providecommand \bibnamefont  [1]{#1}%
\providecommand \bibfnamefont [1]{#1}%
\providecommand \citenamefont [1]{#1}%
\providecommand \href@noop [0]{\@secondoftwo}%
\providecommand \href [0]{\begingroup \@sanitize@url \@href}%
\providecommand \@href[1]{\@@startlink{#1}\@@href}%
\providecommand \@@href[1]{\endgroup#1\@@endlink}%
\providecommand \@sanitize@url [0]{\catcode `\\12\catcode `\$12\catcode
  `\&12\catcode `\#12\catcode `\^12\catcode `\_12\catcode `\%12\relax}%
\providecommand \@@startlink[1]{}%
\providecommand \@@endlink[0]{}%
\providecommand \url  [0]{\begingroup\@sanitize@url \@url }%
\providecommand \@url [1]{\endgroup\@href {#1}{\urlprefix }}%
\providecommand \urlprefix  [0]{URL }%
\providecommand \Eprint [0]{\href }%
\providecommand \doibase [0]{http://dx.doi.org/}%
\providecommand \selectlanguage [0]{\@gobble}%
\providecommand \bibinfo  [0]{\@secondoftwo}%
\providecommand \bibfield  [0]{\@secondoftwo}%
\providecommand \translation [1]{[#1]}%
\providecommand \BibitemOpen [0]{}%
\providecommand \bibitemStop [0]{}%
\providecommand \bibitemNoStop [0]{.\EOS\space}%
\providecommand \EOS [0]{\spacefactor3000\relax}%
\providecommand \BibitemShut  [1]{\csname bibitem#1\endcsname}%
\let\auto@bib@innerbib\@empty
%</preamble>
\bibitem [{\citenamefont {Wen}(2016)}]{Wen16}%
  \BibitemOpen
  \bibfield  {author} {\bibinfo {author} {\bibfnamefont {X.-G.}\ \bibnamefont
  {Wen}},\ }\href@noop {} {\  (\bibinfo {year} {2016})},\ \Eprint
  {http://arxiv.org/abs/1610.03911} {arXiv:1610.03911 [cond-mat.str-el]}
  \BibitemShut {NoStop}%
\bibitem [{\citenamefont {Hasan}\ and\ \citenamefont {Kane}(2010)}]{Hasan10}%
  \BibitemOpen
  \bibfield  {author} {\bibinfo {author} {\bibfnamefont {M.~Z.}\ \bibnamefont
  {Hasan}}\ and\ \bibinfo {author} {\bibfnamefont {C.~L.}\ \bibnamefont
  {Kane}},\ }\href {\doibase 10.1103/RevModPhys.82.3045} {\bibfield  {journal}
  {\bibinfo  {journal} {Rev. Mod. Phys.}\ }\textbf {\bibinfo {volume} {82}},\
  \bibinfo {pages} {3045} (\bibinfo {year} {2010})}\BibitemShut {NoStop}%
\bibitem [{\citenamefont {Gu}\ and\ \citenamefont {Wen}(2009)}]{Gu09}%
  \BibitemOpen
  \bibfield  {author} {\bibinfo {author} {\bibfnamefont {Z.-C.}\ \bibnamefont
  {Gu}}\ and\ \bibinfo {author} {\bibfnamefont {X.-G.}\ \bibnamefont {Wen}},\
  }\href {\doibase 10.1103/PhysRevB.80.155131} {\bibfield  {journal} {\bibinfo
  {journal} {Phys. Rev. B}\ }\textbf {\bibinfo {volume} {80}},\ \bibinfo
  {pages} {155131} (\bibinfo {year} {2009})}\BibitemShut {NoStop}%
\bibitem [{\citenamefont {Pollmann}\ \emph {et~al.}(2012)\citenamefont
  {Pollmann}, \citenamefont {Berg}, \citenamefont {Turner},\ and\ \citenamefont
  {Oshikawa}}]{Pollmann12}%
  \BibitemOpen
  \bibfield  {author} {\bibinfo {author} {\bibfnamefont {F.}~\bibnamefont
  {Pollmann}}, \bibinfo {author} {\bibfnamefont {E.}~\bibnamefont {Berg}},
  \bibinfo {author} {\bibfnamefont {A.~M.}\ \bibnamefont {Turner}}, \ and\
  \bibinfo {author} {\bibfnamefont {M.}~\bibnamefont {Oshikawa}},\ }\href
  {\doibase 10.1103/PhysRevB.85.075125} {\bibfield  {journal} {\bibinfo
  {journal} {Phys. Rev. B}\ }\textbf {\bibinfo {volume} {85}},\ \bibinfo
  {pages} {075125} (\bibinfo {year} {2012})}\BibitemShut {NoStop}%
\bibitem [{\citenamefont {Schnyder}\ \emph {et~al.}(2008)\citenamefont
  {Schnyder}, \citenamefont {Ryu}, \citenamefont {Furusaki},\ and\
  \citenamefont {Ludwig}}]{Schnyder08}%
  \BibitemOpen
  \bibfield  {author} {\bibinfo {author} {\bibfnamefont {A.~P.}\ \bibnamefont
  {Schnyder}}, \bibinfo {author} {\bibfnamefont {S.}~\bibnamefont {Ryu}},
  \bibinfo {author} {\bibfnamefont {A.}~\bibnamefont {Furusaki}}, \ and\
  \bibinfo {author} {\bibfnamefont {A.~W.~W.}\ \bibnamefont {Ludwig}},\ }\href
  {\doibase 10.1103/PhysRevB.78.195125} {\bibfield  {journal} {\bibinfo
  {journal} {Phys. Rev. B}\ }\textbf {\bibinfo {volume} {78}},\ \bibinfo
  {pages} {195125} (\bibinfo {year} {2008})}\BibitemShut {NoStop}%
\bibitem [{\citenamefont {Kitaev}(2009)}]{Kitaev09}%
  \BibitemOpen
  \bibfield  {author} {\bibinfo {author} {\bibfnamefont {A.}~\bibnamefont
  {Kitaev}},\ }\href {\doibase 10.1063/1.3149495} {\bibfield  {journal}
  {\bibinfo  {journal} {AIP Conference Proceedings}\ }\textbf {\bibinfo
  {volume} {1134}},\ \bibinfo {pages} {22} (\bibinfo {year}
  {2009})}\BibitemShut {NoStop}%
\bibitem [{\citenamefont {Ryu}\ \emph {et~al.}(2010)\citenamefont {Ryu},
  \citenamefont {Schnyder}, \citenamefont {Furusaki},\ and\ \citenamefont
  {Ludwig}}]{Ryu10}%
  \BibitemOpen
  \bibfield  {author} {\bibinfo {author} {\bibfnamefont {S.}~\bibnamefont
  {Ryu}}, \bibinfo {author} {\bibfnamefont {A.~P.}\ \bibnamefont {Schnyder}},
  \bibinfo {author} {\bibfnamefont {A.}~\bibnamefont {Furusaki}}, \ and\
  \bibinfo {author} {\bibfnamefont {A.~W.~W.}\ \bibnamefont {Ludwig}},\ }\href
  {http://stacks.iop.org/1367-2630/12/i=6/a=065010} {\bibfield  {journal}
  {\bibinfo  {journal} {New Journal of Physics}\ }\textbf {\bibinfo {volume}
  {12}},\ \bibinfo {pages} {065010} (\bibinfo {year} {2010})}\BibitemShut
  {NoStop}%
\bibitem [{\citenamefont {Fidkowski}\ and\ \citenamefont
  {Kitaev}(2011)}]{Fidkowski11class}%
  \BibitemOpen
  \bibfield  {author} {\bibinfo {author} {\bibfnamefont {L.}~\bibnamefont
  {Fidkowski}}\ and\ \bibinfo {author} {\bibfnamefont {A.}~\bibnamefont
  {Kitaev}},\ }\href {\doibase 10.1103/PhysRevB.83.075103} {\bibfield
  {journal} {\bibinfo  {journal} {Phys. Rev. B}\ }\textbf {\bibinfo {volume}
  {83}},\ \bibinfo {pages} {075103} (\bibinfo {year} {2011})}\BibitemShut
  {NoStop}%
\bibitem [{\citenamefont {Turner}\ \emph {et~al.}(2011)\citenamefont {Turner},
  \citenamefont {Pollmann},\ and\ \citenamefont {Berg}}]{Turner11class}%
  \BibitemOpen
  \bibfield  {author} {\bibinfo {author} {\bibfnamefont {A.~M.}\ \bibnamefont
  {Turner}}, \bibinfo {author} {\bibfnamefont {F.}~\bibnamefont {Pollmann}}, \
  and\ \bibinfo {author} {\bibfnamefont {E.}~\bibnamefont {Berg}},\ }\href
  {\doibase 10.1103/PhysRevB.83.075102} {\bibfield  {journal} {\bibinfo
  {journal} {Phys. Rev. B}\ }\textbf {\bibinfo {volume} {83}},\ \bibinfo
  {pages} {075102} (\bibinfo {year} {2011})}\BibitemShut {NoStop}%
\bibitem [{\citenamefont {Chen}\ \emph {et~al.}(2011)\citenamefont {Chen},
  \citenamefont {Gu},\ and\ \citenamefont {Wen}}]{Chen10}%
  \BibitemOpen
  \bibfield  {author} {\bibinfo {author} {\bibfnamefont {X.}~\bibnamefont
  {Chen}}, \bibinfo {author} {\bibfnamefont {Z.-C.}\ \bibnamefont {Gu}}, \ and\
  \bibinfo {author} {\bibfnamefont {X.-G.}\ \bibnamefont {Wen}},\ }\href
  {\doibase 10.1103/PhysRevB.83.035107} {\bibfield  {journal} {\bibinfo
  {journal} {Phys. Rev. B}\ }\textbf {\bibinfo {volume} {83}},\ \bibinfo
  {pages} {035107} (\bibinfo {year} {2011})}\BibitemShut {NoStop}%
\bibitem [{\citenamefont {{Schuch, N. and P\'erez-Garc\'ia, D. and Cirac, J.
  I.}}(2011)}]{Schuch11}%
  \BibitemOpen
  \bibfield  {author} {\bibinfo {author} {\bibnamefont {{Schuch, N. and
  P\'erez-Garc\'ia, D. and Cirac, J. I.}}},\ }\href {\doibase
  10.1103/PhysRevB.84.165139} {\bibfield  {journal} {\bibinfo  {journal} {Phys.
  Rev. B}\ }\textbf {\bibinfo {volume} {84}},\ \bibinfo {pages} {165139}
  (\bibinfo {year} {2011})}\BibitemShut {NoStop}%
\bibitem [{\citenamefont {Kestner}\ \emph {et~al.}(2011)\citenamefont
  {Kestner}, \citenamefont {Wang}, \citenamefont {Sau},\ and\ \citenamefont
  {Das~Sarma}}]{Kestner11}%
  \BibitemOpen
  \bibfield  {author} {\bibinfo {author} {\bibfnamefont {J.~P.}\ \bibnamefont
  {Kestner}}, \bibinfo {author} {\bibfnamefont {B.}~\bibnamefont {Wang}},
  \bibinfo {author} {\bibfnamefont {J.~D.}\ \bibnamefont {Sau}}, \ and\
  \bibinfo {author} {\bibfnamefont {S.}~\bibnamefont {Das~Sarma}},\ }\href
  {\doibase 10.1103/PhysRevB.83.174409} {\bibfield  {journal} {\bibinfo
  {journal} {Phys. Rev. B}\ }\textbf {\bibinfo {volume} {83}},\ \bibinfo
  {pages} {174409} (\bibinfo {year} {2011})}\BibitemShut {NoStop}%
\bibitem [{\citenamefont {Cheng}\ and\ \citenamefont {Tu}(2011)}]{Cheng11}%
  \BibitemOpen
  \bibfield  {author} {\bibinfo {author} {\bibfnamefont {M.}~\bibnamefont
  {Cheng}}\ and\ \bibinfo {author} {\bibfnamefont {H.-H.}\ \bibnamefont {Tu}},\
  }\href {\doibase 10.1103/PhysRevB.84.094503} {\bibfield  {journal} {\bibinfo
  {journal} {Phys. Rev. B}\ }\textbf {\bibinfo {volume} {84}},\ \bibinfo
  {pages} {094503} (\bibinfo {year} {2011})}\BibitemShut {NoStop}%
\bibitem [{\citenamefont {Fidkowski}\ \emph {et~al.}(2011)\citenamefont
  {Fidkowski}, \citenamefont {Lutchyn}, \citenamefont {Nayak},\ and\
  \citenamefont {Fisher}}]{Fidkowski11longrange}%
  \BibitemOpen
  \bibfield  {author} {\bibinfo {author} {\bibfnamefont {L.}~\bibnamefont
  {Fidkowski}}, \bibinfo {author} {\bibfnamefont {R.~M.}\ \bibnamefont
  {Lutchyn}}, \bibinfo {author} {\bibfnamefont {C.}~\bibnamefont {Nayak}}, \
  and\ \bibinfo {author} {\bibfnamefont {M.~P.~A.}\ \bibnamefont {Fisher}},\
  }\href {\doibase 10.1103/PhysRevB.84.195436} {\bibfield  {journal} {\bibinfo
  {journal} {Phys. Rev. B}\ }\textbf {\bibinfo {volume} {84}},\ \bibinfo
  {pages} {195436} (\bibinfo {year} {2011})}\BibitemShut {NoStop}%
\bibitem [{\citenamefont {Sau}\ \emph {et~al.}(2011)\citenamefont {Sau},
  \citenamefont {Halperin}, \citenamefont {Flensberg},\ and\ \citenamefont
  {Das~Sarma}}]{Sau11}%
  \BibitemOpen
  \bibfield  {author} {\bibinfo {author} {\bibfnamefont {J.~D.}\ \bibnamefont
  {Sau}}, \bibinfo {author} {\bibfnamefont {B.~I.}\ \bibnamefont {Halperin}},
  \bibinfo {author} {\bibfnamefont {K.}~\bibnamefont {Flensberg}}, \ and\
  \bibinfo {author} {\bibfnamefont {S.}~\bibnamefont {Das~Sarma}},\ }\href
  {\doibase 10.1103/PhysRevB.84.144509} {\bibfield  {journal} {\bibinfo
  {journal} {Phys. Rev. B}\ }\textbf {\bibinfo {volume} {84}},\ \bibinfo
  {pages} {144509} (\bibinfo {year} {2011})}\BibitemShut {NoStop}%
\bibitem [{\citenamefont {Kraus}\ \emph {et~al.}(2013)\citenamefont {Kraus},
  \citenamefont {Dalmonte}, \citenamefont {Baranov}, \citenamefont
  {L\"auchli},\ and\ \citenamefont {Zoller}}]{Kraus13}%
  \BibitemOpen
  \bibfield  {author} {\bibinfo {author} {\bibfnamefont {C.~V.}\ \bibnamefont
  {Kraus}}, \bibinfo {author} {\bibfnamefont {M.}~\bibnamefont {Dalmonte}},
  \bibinfo {author} {\bibfnamefont {M.~A.}\ \bibnamefont {Baranov}}, \bibinfo
  {author} {\bibfnamefont {A.~M.}\ \bibnamefont {L\"auchli}}, \ and\ \bibinfo
  {author} {\bibfnamefont {P.}~\bibnamefont {Zoller}},\ }\href {\doibase
  10.1103/PhysRevLett.111.173004} {\bibfield  {journal} {\bibinfo  {journal}
  {Phys. Rev. Lett.}\ }\textbf {\bibinfo {volume} {111}},\ \bibinfo {pages}
  {173004} (\bibinfo {year} {2013})}\BibitemShut {NoStop}%
\bibitem [{\citenamefont {Keselman}\ and\ \citenamefont
  {Berg}(2015)}]{Keselman15}%
  \BibitemOpen
  \bibfield  {author} {\bibinfo {author} {\bibfnamefont {A.}~\bibnamefont
  {Keselman}}\ and\ \bibinfo {author} {\bibfnamefont {E.}~\bibnamefont
  {Berg}},\ }\href {\doibase 10.1103/PhysRevB.91.235309} {\bibfield  {journal}
  {\bibinfo  {journal} {Phys. Rev. B}\ }\textbf {\bibinfo {volume} {91}},\
  \bibinfo {pages} {235309} (\bibinfo {year} {2015})}\BibitemShut {NoStop}%
\bibitem [{\citenamefont {Iemini}\ \emph {et~al.}(2015)\citenamefont {Iemini},
  \citenamefont {Mazza}, \citenamefont {Rossini}, \citenamefont {Fazio},\ and\
  \citenamefont {Diehl}}]{Iemini15}%
  \BibitemOpen
  \bibfield  {author} {\bibinfo {author} {\bibfnamefont {F.}~\bibnamefont
  {Iemini}}, \bibinfo {author} {\bibfnamefont {L.}~\bibnamefont {Mazza}},
  \bibinfo {author} {\bibfnamefont {D.}~\bibnamefont {Rossini}}, \bibinfo
  {author} {\bibfnamefont {R.}~\bibnamefont {Fazio}}, \ and\ \bibinfo {author}
  {\bibfnamefont {S.}~\bibnamefont {Diehl}},\ }\href {\doibase
  10.1103/PhysRevLett.115.156402} {\bibfield  {journal} {\bibinfo  {journal}
  {Phys. Rev. Lett.}\ }\textbf {\bibinfo {volume} {115}},\ \bibinfo {pages}
  {156402} (\bibinfo {year} {2015})}\BibitemShut {NoStop}%
\bibitem [{\citenamefont {Lang}\ and\ \citenamefont
  {B\"uchler}(2015)}]{Lang15}%
  \BibitemOpen
  \bibfield  {author} {\bibinfo {author} {\bibfnamefont {N.}~\bibnamefont
  {Lang}}\ and\ \bibinfo {author} {\bibfnamefont {H.~P.}\ \bibnamefont
  {B\"uchler}},\ }\href {\doibase 10.1103/PhysRevB.92.041118} {\bibfield
  {journal} {\bibinfo  {journal} {Phys. Rev. B}\ }\textbf {\bibinfo {volume}
  {92}},\ \bibinfo {pages} {041118} (\bibinfo {year} {2015})}\BibitemShut
  {NoStop}%
\bibitem [{\citenamefont {Montorsi}\ \emph {et~al.}(2017)\citenamefont
  {Montorsi}, \citenamefont {Dolcini}, \citenamefont {Iotti},\ and\
  \citenamefont {Rossi}}]{Montorsi17}%
  \BibitemOpen
  \bibfield  {author} {\bibinfo {author} {\bibfnamefont {A.}~\bibnamefont
  {Montorsi}}, \bibinfo {author} {\bibfnamefont {F.}~\bibnamefont {Dolcini}},
  \bibinfo {author} {\bibfnamefont {R.~C.}\ \bibnamefont {Iotti}}, \ and\
  \bibinfo {author} {\bibfnamefont {F.}~\bibnamefont {Rossi}},\ }\href
  {\doibase 10.1103/PhysRevB.95.245108} {\bibfield  {journal} {\bibinfo
  {journal} {Phys. Rev. B}\ }\textbf {\bibinfo {volume} {95}},\ \bibinfo
  {pages} {245108} (\bibinfo {year} {2017})}\BibitemShut {NoStop}%
\bibitem [{\citenamefont {Ruhman}\ and\ \citenamefont
  {Altman}(2017)}]{Ruhman17}%
  \BibitemOpen
  \bibfield  {author} {\bibinfo {author} {\bibfnamefont {J.}~\bibnamefont
  {Ruhman}}\ and\ \bibinfo {author} {\bibfnamefont {E.}~\bibnamefont
  {Altman}},\ }\href {\doibase 10.1103/PhysRevB.96.085133} {\bibfield
  {journal} {\bibinfo  {journal} {Phys. Rev. B}\ }\textbf {\bibinfo {volume}
  {96}},\ \bibinfo {pages} {085133} (\bibinfo {year} {2017})}\BibitemShut
  {NoStop}%
\bibitem [{\citenamefont {{Scaffidi}}\ \emph {et~al.}(2017)\citenamefont
  {{Scaffidi}}, \citenamefont {{Parker}},\ and\ \citenamefont
  {{Vasseur}}}]{Scaffidi17_preprint}%
  \BibitemOpen
  \bibfield  {author} {\bibinfo {author} {\bibfnamefont {T.}~\bibnamefont
  {{Scaffidi}}}, \bibinfo {author} {\bibfnamefont {D.~E.}\ \bibnamefont
  {{Parker}}}, \ and\ \bibinfo {author} {\bibfnamefont {R.}~\bibnamefont
  {{Vasseur}}},\ }\href@noop {} {\bibfield  {journal} {\bibinfo  {journal}
  {ArXiv e-prints}\ } (\bibinfo {year} {2017})},\ \Eprint
  {http://arxiv.org/abs/1705.01557} {arXiv:1705.01557 [cond-mat.str-el]}
  \BibitemShut {NoStop}%
\bibitem [{\citenamefont {{Jiang}}\ \emph {et~al.}(2017)\citenamefont
  {{Jiang}}, \citenamefont {{Li}}, \citenamefont {{Seidel}},\ and\
  \citenamefont {{Lee}}}]{Jiang17_preprint}%
  \BibitemOpen
  \bibfield  {author} {\bibinfo {author} {\bibfnamefont {H.-C.}\ \bibnamefont
  {{Jiang}}}, \bibinfo {author} {\bibfnamefont {Z.-X.}\ \bibnamefont {{Li}}},
  \bibinfo {author} {\bibfnamefont {A.}~\bibnamefont {{Seidel}}}, \ and\
  \bibinfo {author} {\bibfnamefont {D.-H.}\ \bibnamefont {{Lee}}},\ }\href@noop
  {} {\bibfield  {journal} {\bibinfo  {journal} {ArXiv e-prints}\ } (\bibinfo
  {year} {2017})},\ \Eprint {http://arxiv.org/abs/1704.02997} {arXiv:1704.02997
  [cond-mat.str-el]} \BibitemShut {NoStop}%
\bibitem [{\citenamefont {{Zhang}}\ and\ \citenamefont
  {{Liu}}(2017)}]{Zhang17}%
  \BibitemOpen
  \bibfield  {author} {\bibinfo {author} {\bibfnamefont {R.-X.}\ \bibnamefont
  {{Zhang}}}\ and\ \bibinfo {author} {\bibfnamefont {C.-X.}\ \bibnamefont
  {{Liu}}},\ }\href@noop {} {\bibfield  {journal} {\bibinfo  {journal} {ArXiv
  e-prints}\ } (\bibinfo {year} {2017})},\ \Eprint
  {http://arxiv.org/abs/1708.01367} {arXiv:1708.01367 [cond-mat.str-el]}
  \BibitemShut {NoStop}%
\bibitem [{\citenamefont {Altland}\ and\ \citenamefont
  {Zirnbauer}(1997)}]{Altland97}%
  \BibitemOpen
  \bibfield  {author} {\bibinfo {author} {\bibfnamefont {A.}~\bibnamefont
  {Altland}}\ and\ \bibinfo {author} {\bibfnamefont {M.~R.}\ \bibnamefont
  {Zirnbauer}},\ }\href {\doibase 10.1103/PhysRevB.55.1142} {\bibfield
  {journal} {\bibinfo  {journal} {Phys. Rev. B}\ }\textbf {\bibinfo {volume}
  {55}},\ \bibinfo {pages} {1142} (\bibinfo {year} {1997})}\BibitemShut
  {NoStop}%
\bibitem [{\citenamefont {Motrunich}\ \emph {et~al.}(2001)\citenamefont
  {Motrunich}, \citenamefont {Damle},\ and\ \citenamefont
  {Huse}}]{Motrunich01}%
  \BibitemOpen
  \bibfield  {author} {\bibinfo {author} {\bibfnamefont {O.}~\bibnamefont
  {Motrunich}}, \bibinfo {author} {\bibfnamefont {K.}~\bibnamefont {Damle}}, \
  and\ \bibinfo {author} {\bibfnamefont {D.~A.}\ \bibnamefont {Huse}},\ }\href
  {\doibase 10.1103/PhysRevB.63.224204} {\bibfield  {journal} {\bibinfo
  {journal} {Phys. Rev. B}\ }\textbf {\bibinfo {volume} {63}},\ \bibinfo
  {pages} {224204} (\bibinfo {year} {2001})}\BibitemShut {NoStop}%
\bibitem [{\citenamefont {Brouwer}\ \emph {et~al.}(1998)\citenamefont
  {Brouwer}, \citenamefont {Mudry}, \citenamefont {Simons},\ and\ \citenamefont
  {Altland}}]{Brouwer98}%
  \BibitemOpen
  \bibfield  {author} {\bibinfo {author} {\bibfnamefont {P.~W.}\ \bibnamefont
  {Brouwer}}, \bibinfo {author} {\bibfnamefont {C.}~\bibnamefont {Mudry}},
  \bibinfo {author} {\bibfnamefont {B.~D.}\ \bibnamefont {Simons}}, \ and\
  \bibinfo {author} {\bibfnamefont {A.}~\bibnamefont {Altland}},\ }\href
  {\doibase 10.1103/PhysRevLett.81.862} {\bibfield  {journal} {\bibinfo
  {journal} {Phys. Rev. Lett.}\ }\textbf {\bibinfo {volume} {81}},\ \bibinfo
  {pages} {862} (\bibinfo {year} {1998})}\BibitemShut {NoStop}%
\bibitem [{\citenamefont {Fulga}\ \emph {et~al.}(2011)\citenamefont {Fulga},
  \citenamefont {Hassler}, \citenamefont {Akhmerov},\ and\ \citenamefont
  {Beenakker}}]{Fulga11}%
  \BibitemOpen
  \bibfield  {author} {\bibinfo {author} {\bibfnamefont {I.~C.}\ \bibnamefont
  {Fulga}}, \bibinfo {author} {\bibfnamefont {F.}~\bibnamefont {Hassler}},
  \bibinfo {author} {\bibfnamefont {A.~R.}\ \bibnamefont {Akhmerov}}, \ and\
  \bibinfo {author} {\bibfnamefont {C.~W.~J.}\ \bibnamefont {Beenakker}},\
  }\href {\doibase 10.1103/PhysRevB.83.155429} {\bibfield  {journal} {\bibinfo
  {journal} {Phys. Rev. B}\ }\textbf {\bibinfo {volume} {83}},\ \bibinfo
  {pages} {155429} (\bibinfo {year} {2011})}\BibitemShut {NoStop}%
\bibitem [{\citenamefont {{DeGottardi}}\ \emph {et~al.}(2013)\citenamefont
  {{DeGottardi}}, \citenamefont {{Thakurathi}}, \citenamefont
  {{Vishveshwara}},\ and\ \citenamefont {{Sen}}}]{deGottardi13}%
  \BibitemOpen
  \bibfield  {author} {\bibinfo {author} {\bibfnamefont {W.}~\bibnamefont
  {{DeGottardi}}}, \bibinfo {author} {\bibfnamefont {M.}~\bibnamefont
  {{Thakurathi}}}, \bibinfo {author} {\bibfnamefont {S.}~\bibnamefont
  {{Vishveshwara}}}, \ and\ \bibinfo {author} {\bibfnamefont {D.}~\bibnamefont
  {{Sen}}},\ }\href {\doibase 10.1103/PhysRevB.88.165111} {\bibfield  {journal}
  {\bibinfo  {journal} {\prb}\ }\textbf {\bibinfo {volume} {88}},\ \bibinfo
  {eid} {165111} (\bibinfo {year} {2013})}\BibitemShut {NoStop}%
\bibitem [{\citenamefont {{Das}}\ and\ \citenamefont {{Satija}}(2014)}]{Das14}%
  \BibitemOpen
  \bibfield  {author} {\bibinfo {author} {\bibfnamefont {S.}~\bibnamefont
  {{Das}}}\ and\ \bibinfo {author} {\bibfnamefont {I.~I.}\ \bibnamefont
  {{Satija}}},\ }\href@noop {} {\bibfield  {journal} {\bibinfo  {journal}
  {ArXiv e-prints}\ } (\bibinfo {year} {2014})},\ \Eprint
  {http://arxiv.org/abs/1409.6139} {arXiv:1409.6139 [cond-mat.mes-hall]}
  \BibitemShut {NoStop}%
\bibitem [{\citenamefont {Mandal}(2015)}]{Mandal15}%
  \BibitemOpen
  \bibfield  {author} {\bibinfo {author} {\bibfnamefont {I.}~\bibnamefont
  {Mandal}},\ }\href
  {http://iopscience.iop.org/article/10.1209/0295-5075/110/67005/meta}
  {\bibfield  {journal} {\bibinfo  {journal} {EPL (Europhysics Letters)}\
  }\textbf {\bibinfo {volume} {110}},\ \bibinfo {pages} {67005} (\bibinfo
  {year} {2015})}\BibitemShut {NoStop}%
\bibitem [{\citenamefont {Verresen}\ \emph {et~al.}(2017)\citenamefont
  {Verresen}, \citenamefont {Moessner},\ and\ \citenamefont
  {Pollmann}}]{Verresen17}%
  \BibitemOpen
  \bibfield  {author} {\bibinfo {author} {\bibfnamefont {R.}~\bibnamefont
  {Verresen}}, \bibinfo {author} {\bibfnamefont {R.}~\bibnamefont {Moessner}},
  \ and\ \bibinfo {author} {\bibfnamefont {F.}~\bibnamefont {Pollmann}},\
  }\href {\doibase 10.1103/PhysRevB.96.165124} {\bibfield  {journal} {\bibinfo
  {journal} {Phys. Rev. B}\ }\textbf {\bibinfo {volume} {96}},\ \bibinfo
  {pages} {165124} (\bibinfo {year} {2017})}\BibitemShut {NoStop}%
\bibitem [{\citenamefont {Niu}\ \emph {et~al.}(2012)\citenamefont {Niu},
  \citenamefont {Chung}, \citenamefont {Hsu}, \citenamefont {Mandal},
  \citenamefont {Raghu},\ and\ \citenamefont {Chakravarty}}]{Niu12}%
  \BibitemOpen
  \bibfield  {author} {\bibinfo {author} {\bibfnamefont {Y.}~\bibnamefont
  {Niu}}, \bibinfo {author} {\bibfnamefont {S.~B.}\ \bibnamefont {Chung}},
  \bibinfo {author} {\bibfnamefont {C.-H.}\ \bibnamefont {Hsu}}, \bibinfo
  {author} {\bibfnamefont {I.}~\bibnamefont {Mandal}}, \bibinfo {author}
  {\bibfnamefont {S.}~\bibnamefont {Raghu}}, \ and\ \bibinfo {author}
  {\bibfnamefont {S.}~\bibnamefont {Chakravarty}},\ }\href {\doibase
  10.1103/PhysRevB.85.035110} {\bibfield  {journal} {\bibinfo  {journal} {Phys.
  Rev. B}\ }\textbf {\bibinfo {volume} {85}},\ \bibinfo {pages} {035110}
  (\bibinfo {year} {2012})}\BibitemShut {NoStop}%
\bibitem [{\citenamefont {Kitaev}(2001)}]{Kitaev01}%
  \BibitemOpen
  \bibfield  {author} {\bibinfo {author} {\bibfnamefont {A.}~\bibnamefont
  {Kitaev}},\ }\href@noop {} {\bibfield  {journal} {\bibinfo  {journal}
  {Physics-Uspekhi}\ }\textbf {\bibinfo {volume} {44}},\ \bibinfo {pages} {131}
  (\bibinfo {year} {2001})}\BibitemShut {NoStop}%
\bibitem [{\citenamefont {{Asb{\'o}th}}\ \emph {et~al.}(2016)\citenamefont
  {{Asb{\'o}th}}, \citenamefont {{Oroszl{\'a}ny}},\ and\ \citenamefont
  {{P{\'a}lyi}}}]{Asboth15}%
  \BibitemOpen
  \bibfield  {author} {\bibinfo {author} {\bibfnamefont {J.~K.}\ \bibnamefont
  {{Asb{\'o}th}}}, \bibinfo {author} {\bibfnamefont {L.}~\bibnamefont
  {{Oroszl{\'a}ny}}}, \ and\ \bibinfo {author} {\bibfnamefont {A.}~\bibnamefont
  {{P{\'a}lyi}}},\ }\href {\doibase 10.1007/978-3-319-25607-8} {\emph {\bibinfo
  {title} {A Short Course on Topological Insulators}}}\ (\bibinfo  {publisher}
  {Springer International Publishing},\ \bibinfo {year} {2016})\BibitemShut
  {NoStop}%
\bibitem [{Note1()}]{Note1}%
  \BibitemOpen
  \bibinfo {note} {One can also imagine a box function, corresponding to $\xi =
  0$.}\BibitemShut {Stop}%
\bibitem [{\citenamefont {Di~Francesco}\ \emph {et~al.}(1997)\citenamefont
  {Di~Francesco}, \citenamefont {Mathieu},\ and\ \citenamefont
  {S\'en\'echal}}]{CFT}%
  \BibitemOpen
  \bibfield  {author} {\bibinfo {author} {\bibfnamefont {P.}~\bibnamefont
  {Di~Francesco}}, \bibinfo {author} {\bibfnamefont {P.}~\bibnamefont
  {Mathieu}}, \ and\ \bibinfo {author} {\bibfnamefont {D.}~\bibnamefont
  {S\'en\'echal}},\ }\href {\doibase 10.1007/978-1-4612-2256-9} {\emph
  {\bibinfo {title} {{Conformal field theory}}}},\ Graduate texts in
  contemporary physics\ (\bibinfo  {publisher} {Springer},\ \bibinfo {year}
  {1997})\BibitemShut {NoStop}%
\bibitem [{sup()}]{suppl}%
  \BibitemOpen
  \href@noop {} {}\bibinfo {note} {See Supplemental Material.}\BibitemShut
  {Stop}%
\bibitem [{\citenamefont {Furuya}\ and\ \citenamefont
  {Oshikawa}(2017)}]{Furuya17}%
  \BibitemOpen
  \bibfield  {author} {\bibinfo {author} {\bibfnamefont {S.~C.}\ \bibnamefont
  {Furuya}}\ and\ \bibinfo {author} {\bibfnamefont {M.}~\bibnamefont
  {Oshikawa}},\ }\href {\doibase 10.1103/PhysRevLett.118.021601} {\bibfield
  {journal} {\bibinfo  {journal} {Phys. Rev. Lett.}\ }\textbf {\bibinfo
  {volume} {118}},\ \bibinfo {pages} {021601} (\bibinfo {year}
  {2017})}\BibitemShut {NoStop}%
\bibitem [{\citenamefont {Zamolodchikov}(1986)}]{Zamolodchikov1986}%
  \BibitemOpen
  \bibfield  {author} {\bibinfo {author} {\bibfnamefont {A.~B.}\ \bibnamefont
  {Zamolodchikov}},\ }\href
  {http://www.jetpletters.ac.ru/ps/1413/article_21504.shtml} {\bibfield
  {journal} {\bibinfo  {journal} {JETP Lett.}\ }\textbf {\bibinfo {volume}
  {43}},\ \bibinfo {pages} {730} (\bibinfo {year} {1986})}\BibitemShut
  {NoStop}%
\bibitem [{Note2()}]{Note2}%
  \BibitemOpen
  \bibinfo {note} {E.g. $f(z) = (z-b-1)(z+|b|+1)$ at $b=0$ is a CFT with $c=1$
  between the trivial phase and the Kitaev chain phase.}\BibitemShut {Stop}%
\bibitem [{\citenamefont {Refael}\ and\ \citenamefont
  {Moore}(2004)}]{Refael04}%
  \BibitemOpen
  \bibfield  {author} {\bibinfo {author} {\bibfnamefont {G.}~\bibnamefont
  {Refael}}\ and\ \bibinfo {author} {\bibfnamefont {J.~E.}\ \bibnamefont
  {Moore}},\ }\href {\doibase 10.1103/PhysRevLett.93.260602} {\bibfield
  {journal} {\bibinfo  {journal} {Phys. Rev. Lett.}\ }\textbf {\bibinfo
  {volume} {93}},\ \bibinfo {pages} {260602} (\bibinfo {year}
  {2004})}\BibitemShut {NoStop}%
\bibitem [{\citenamefont {Laflorencie}(2005)}]{Laflorencie05}%
  \BibitemOpen
  \bibfield  {author} {\bibinfo {author} {\bibfnamefont {N.}~\bibnamefont
  {Laflorencie}},\ }\href {\doibase 10.1103/PhysRevB.72.140408} {\bibfield
  {journal} {\bibinfo  {journal} {Phys. Rev. B}\ }\textbf {\bibinfo {volume}
  {72}},\ \bibinfo {pages} {140408} (\bibinfo {year} {2005})}\BibitemShut
  {NoStop}%
\bibitem [{\citenamefont {Fidkowski}\ and\ \citenamefont
  {Kitaev}(2010)}]{Fidkowski10interaction}%
  \BibitemOpen
  \bibfield  {author} {\bibinfo {author} {\bibfnamefont {L.}~\bibnamefont
  {Fidkowski}}\ and\ \bibinfo {author} {\bibfnamefont {A.}~\bibnamefont
  {Kitaev}},\ }\href {\doibase 10.1103/PhysRevB.81.134509} {\bibfield
  {journal} {\bibinfo  {journal} {Phys. Rev. B}\ }\textbf {\bibinfo {volume}
  {81}},\ \bibinfo {pages} {134509} (\bibinfo {year} {2010})}\BibitemShut
  {NoStop}%
\bibitem [{\citenamefont {White}(1992)}]{White92}%
  \BibitemOpen
  \bibfield  {author} {\bibinfo {author} {\bibfnamefont {S.~R.}\ \bibnamefont
  {White}},\ }\href {\doibase 10.1103/PhysRevLett.69.2863} {\bibfield
  {journal} {\bibinfo  {journal} {Phys. Rev. Lett.}\ }\textbf {\bibinfo
  {volume} {69}},\ \bibinfo {pages} {2863} (\bibinfo {year}
  {1992})}\BibitemShut {NoStop}%
\bibitem [{\citenamefont {Calabrese}\ and\ \citenamefont
  {Cardy}(2004)}]{Calabrese04}%
  \BibitemOpen
  \bibfield  {author} {\bibinfo {author} {\bibfnamefont {P.}~\bibnamefont
  {Calabrese}}\ and\ \bibinfo {author} {\bibfnamefont {J.}~\bibnamefont
  {Cardy}},\ }\href {\doibase 10.1088/1742-5468/2004/06/P06002} {\bibfield
  {journal} {\bibinfo  {journal} {J. Stat. Mech.}\ }\textbf {\bibinfo {volume}
  {2004}},\ \bibinfo {pages} {P06002} (\bibinfo {year} {2004})}\BibitemShut
  {NoStop}%
\bibitem [{\citenamefont {Pollmann}\ \emph {et~al.}(2010)\citenamefont
  {Pollmann}, \citenamefont {Turner}, \citenamefont {Berg},\ and\ \citenamefont
  {Oshikawa}}]{Pollmann10}%
  \BibitemOpen
  \bibfield  {author} {\bibinfo {author} {\bibfnamefont {F.}~\bibnamefont
  {Pollmann}}, \bibinfo {author} {\bibfnamefont {A.~M.}\ \bibnamefont
  {Turner}}, \bibinfo {author} {\bibfnamefont {E.}~\bibnamefont {Berg}}, \ and\
  \bibinfo {author} {\bibfnamefont {M.}~\bibnamefont {Oshikawa}},\ }\href
  {\doibase 10.1103/PhysRevB.81.064439} {\bibfield  {journal} {\bibinfo
  {journal} {Phys. Rev. B}\ }\textbf {\bibinfo {volume} {81}},\ \bibinfo
  {pages} {064439} (\bibinfo {year} {2010})}\BibitemShut {NoStop}%
\bibitem [{\citenamefont {Jones}\ and\ \citenamefont
  {Verresen}(2018)}]{Jonesfuture}%
  \BibitemOpen
  \bibfield  {author} {\bibinfo {author} {\bibfnamefont {N.~G.}\ \bibnamefont
  {Jones}}\ and\ \bibinfo {author} {\bibfnamefont {R.}~\bibnamefont
  {Verresen}},\ }\href@noop {} {} (\bibinfo {year} {2018}),\ \bibinfo {note}
  {to appear}\BibitemShut {NoStop}%
\bibitem [{Note3()}]{Note3}%
  \BibitemOpen
  \bibinfo {note} {In principle we should add `per central charge', but only
  the smallest allowed one will be stable.}\BibitemShut {Stop}%
\bibitem [{\citenamefont {{Vidal}}(2007)}]{Vidal07}%
  \BibitemOpen
  \bibfield  {author} {\bibinfo {author} {\bibfnamefont {G.}~\bibnamefont
  {{Vidal}}},\ }\href {\doibase 10.1103/PhysRevLett.99.220405} {\bibfield
  {journal} {\bibinfo  {journal} {Physical Review Letters}\ }\textbf {\bibinfo
  {volume} {99}},\ \bibinfo {eid} {220405} (\bibinfo {year}
  {2007})}\BibitemShut {NoStop}%
\bibitem [{\citenamefont {Cirac}\ and\ \citenamefont {Sierra}(2010)}]{Cirac10}%
  \BibitemOpen
  \bibfield  {author} {\bibinfo {author} {\bibfnamefont {J.~I.}\ \bibnamefont
  {Cirac}}\ and\ \bibinfo {author} {\bibfnamefont {G.}~\bibnamefont {Sierra}},\
  }\href {\doibase 10.1103/PhysRevB.81.104431} {\bibfield  {journal} {\bibinfo
  {journal} {Phys. Rev. B}\ }\textbf {\bibinfo {volume} {81}},\ \bibinfo
  {pages} {104431} (\bibinfo {year} {2010})}\BibitemShut {NoStop}%
\end{thebibliography}%
%\bibliography{bibo}

\pagebreak

\widetext
\ifx\targetformat\undefined
\begin{center}
	\textbf{\large Supplemental Materials: ``Topology and edge modes in quantum critical chains''}
\end{center}

\setcounter{equation}{0}
\setcounter{figure}{0}
\setcounter{table}{0}
\makeatletter
\renewcommand{\theequation}{S\arabic{equation}}
\renewcommand{\thefigure}{S\arabic{figure}}
\renewcommand{\bibnumfmt}[1]{[S#1]}
\renewcommand{\citenumfont}[1]{S#1}

\else
\appendix
\fi

\section{Solving the translation invariant model with periodic boundary conditions} \label{app:solution}

Let our Hamiltonian be given by $H = \sum t_\alpha H_\alpha$. Define $f(k) = \sum_\alpha t_\alpha e^{ik\alpha}$ and $\varepsilon_k,\varphi_k \in \mathbb R$ such that $f(k) = \varepsilon_k e^{i\varphi_k}$. One might choose to take $\varepsilon_k \geq 0$, but we do not require this. We now prove that a Bogoliubov rotation with angle $\varphi_k$ diagonalizes the Hamiltonian, with single-particle spectrum $\varepsilon_k$, i.e.:
\begin{equation} \label{ham_solved}
H = \sum_k \varepsilon_k \; \left( \frac{1}{2} - d_k^\dagger d_k \right) \; \qquad \textrm{where } \left(\begin{array}{c} d_k \\ d_{-k}^\dagger \end{array}\right) = \exp \left(i \varphi_k \frac{\sigma^x}{2}\right) \; \left(\begin{array}{c} c_k \\ c_{-k}^\dagger \end{array}\right) \textrm{ with } c_k = \frac{1}{\sqrt{L}} \sum e^{-ikn} c_n \; .
\end{equation}
To see this, first consider the $\alpha$-chain $H_\alpha = \frac{i}{2} \sum \tilde \gamma_n \gamma_{n+\alpha}$. In terms of the Fourier modes, one obtains
\begin{equation}
H_\alpha = - \frac{1}{2}\sum_k \left( c_k^\dagger \quad c_{-k} \right) \; H_{\alpha,k} \;  \left(\begin{array}{c} c_k \\ c_{-k}^\dagger \end{array}\right) \qquad \textrm{ where } H_{\alpha,k} = \cos(-k\alpha) \; \sigma^z + \sin(-k \alpha ) \; \sigma^y \; .
\end{equation}
So for our original Hamiltonian $H = \sum_\alpha t_\alpha \; H_\alpha$ we thus obtain
\begin{equation}
H = - \frac{1}{2}\sum_k \left( c_k^\dagger \quad c_{-k} \right) \; H_k \;  \left(\begin{array}{c} c_k \\ c_{-k}^\dagger \end{array}\right) \qquad \textrm{ where } H_k = \varepsilon_k \; \left( \cos(-\varphi_k) \; \sigma^z + \sin(-\varphi_k) \; \sigma^y \right) \; .
\end{equation}
We can interpret $H_k$ as two-dimensional vector which we can align with the $\sigma^z$ axis by rotating it over an angle $-\varphi_k$ around the $\sigma^x$ axis. This is implemented by $U(\vartheta) = \exp{\left( -i \vartheta S^x\right)} = \exp{\left(- i \vartheta \sigma^x /2 \right)}$, such that $H_k = \varepsilon_k \; U(\varphi_k) \; \sigma^z \; U(-\varphi_k)$. Hence,
\begin{align}
H &= - \frac{1}{2}\sum_k \varepsilon_k \; \left( d_k^\dagger \quad d_{-k} \right) \; \sigma^z \;  \left(\begin{array}{c} d_k \\ d_{-k}^\dagger \end{array}\right) = - \sum_k \varepsilon_k \; d_k^\dagger d_k + \frac{1}{2} \sum_k \varepsilon_k \; .
\end{align}

\ifx\targetformat\undefined
\section{Majorana edge modes: details for proof of Theorem 1}
\else
\section{Majorana edge modes: details for proof of Theorem \ref{thm}}
\fi
\label{app:proof}

First we treat the case $\omega> 0$, i.e. the zeros strictly within the unit disk outnumber the order of the pole at the origin ($N_z>N_p$). We consider $H = \frac{i}{2} \sum_{\alpha=-\infty}^{+\infty} t_\alpha \; \left(\sum_{n = 1}^\infty \tilde \gamma_n \gamma_{n+\alpha} \right)$ on the half-infinite chain. Let $\{ z_i \}$ denote the $\omega$ largest zeros strictly within the unit disk. We now show that for each such $z_i$, we can construct a real Majorana edge mode on the left edge:
\begin{equation}
\gamma_\textrm{left}^{(i)} = \sum_{n\geq 1} b_n^{(i)} \; \gamma_n\; .
\end{equation}
Note that if the coefficients $b_n^{(i)}$ are real, this is indeed Hermitian and commutes with $T$. The remaining requirements are hence that, firstly, $\gamma_\textrm{left}^{(i)}$ commutes with $H$, secondly, that $|b_n^{(i)}| \sim |z_i|^n$, and lastly, that the different modes anti-commute, i.e. that $\left\{\gamma_\textrm{left}^{(i)},\gamma_\textrm{left}^{(j)} \right\} = 2\delta_{ij}$. However, if the latter is not satisfied, this can be remedied by noting that $\{\cdot,\cdot \}$ defines an inner product on the space of zero modes and hence one can apply the Gramm-Schmidt process. Note that one should do this in order of ascending correlation lengths, so as to not affect the dominant part of the spatial decay. In conclusion, it is sufficient to fix one of the $z_i$ and show that we can find real-valued $b_n^{(i)}$ such that $[\gamma_\textrm{left}^{(i)},H] = 0$ and $|b_n^{(i)}| \sim |z_i|^n$ (the latter implying a localization length $\xi_i = -\frac{1}{\ln |z_i|}$). The resulting edge modes need to be linearly independent, which is automatic for distinct zeros (due to the distinct asymptotic forms) but we will have to take care when there are degenerate zeros.

A straight-forward calculation shows that $[\gamma_\textrm{left}^{(i)},H] = -i\sum_{n\geq 1} \mathcal C_n \tilde \gamma_n$ with
\begin{equation}
\mathcal C_n = \sum_{a\geq 1} b_a^{(i)} \; t_{a-n} 
\end{equation}
giving us an infinite number of constraints $\{\mathcal C_n = 0\}$. Let us first consider the case $N_p = 0$, then $t_{\alpha < 0} = 0$, which means that each $\mathcal C_n$ contains \emph{all} the coefficients of $f(z)$. Concretely, this means that upon taking $b_a^{(i)} = z_i^{a-1}$, we have $\mathcal C_n = z_i^{n-1} f(z_i)$ and hence all constraints are trivially satisfied! If $z_i$ is real, this defines a real Majorana mode, which is normalizable since $\left\{\gamma_\textrm{left}^{(i)},\gamma_\textrm{left}^{(i)} \right\} =  \sum_{n,m = 1}^\infty  z_i^{n-1} z_i^{m-1} 2\delta_{nm} = \frac{2}{1-z_i^2} \neq 0$. If $z_i$ is complex, we can choose $b_n^{(i)} = z_i^{n-1} \pm \overline z_i^{n-1}$ (which are indeed also solutions due to the Hermiticity of $\gamma_n$). Hence if $z_i$ is complex, we obtain \emph{two} solutions, consistent with $z_i$ and $\overline z_i$ being distinct zeros of $f(z)$. Another subtlety arises when $z_i$ is degenerate. Suppose $z_i$ has an $m$-fold multiplicity (i.e. there are $i_1, \cdots, i_m$ such that $z_{i_1} = \cdots = z_{i_m}$), then $b_a^{(i_{l})} = \frac{\mathrm d^{l-1} z^{a-1}}{\mathrm d z^{l-1}} \big|_{z = z_i} = \frac{(a-1)!}{(a-l)!}z_i^{a-l} $ (with $l = 1,\cdots,m$) define $m$ solutions since then $\mathcal C_n = \left( z^{n-1} \frac{\mathrm d^{l-1} f(z)}{\mathrm d z^{l-1}} \right)\Big|_{z = z_i} = 0 $. Note that these solutions are linearly independent, yet they all have the same localization length.

We now consider the case $N_p > 0$. Let $\{ \tilde z_s\}$ denote the $N_p$ \emph{smallest} zeros of $f(z)$. Since $\omega = N_z - N_p > 0$ by assumption, we know that $z_i \notin \{ \tilde z_s\}$. We then consider the ansatz
\begin{equation}
b_a^{(i)} = z_i^a + \sum_{s = 1}^{N_p} \lambda_s^{(i)} \tilde z_s^a \; .
\end{equation}
Note that before we had to take the exponent of $z_i$ to be $a-1$ instead of $a$ to ensure normalizability in case $z_i = 0$, but when $N_p > 0$ we know that $z_i \neq 0$ (otherwise it could be substracted from the pole). Since $t_{\alpha < -N_p} = 0$, we similarly have that $\mathcal C_{n>N_p} =0$ is trivially satisfied by virtue of $z_i$ and $\{\tilde z_s\}$ being zeros of $f(z)$. The remaining $N_p$ conditions are equivalent to a problem of the type $A \lambda = b$, where the $N_p \times N_p$ matrix $A_{ns} = \left( \sum_{a\geq 1} t_{a-n} \; \tilde z_s^a \right)$. We now show that if the zeros are not degenerate, $A$ is invertible. Indeed: by virtue of $f(z_i)=0$ one can rewrite $A_{ns}= - \sum_{a=n}^{N_p} t_{-a} \tilde z_s^{n-a}$ (using that $\tilde z_s \neq 0$ since $N_p > 0$), which by simple row reduction can be reduced to the Vandermonde matrix associated to $\{\tilde z_s\}$ with determinant $\prod (\tilde z_s-\tilde z_{s'}) \neq 0$. Hence there is a unique solution for $\{\lambda_i\}$. If there \emph{is} a degeneracy in $\{ \tilde z_s \}$, one can repeat the trick encountered in the case $N_p = 0$ by working with the derivatives instead. Similarly, if one of the zeros is complex, one can take the real and imaginary combinations of the above solution. In the special case that $\overline z_i \in \{ \tilde z_s\}$, these two solutions are not linearly independent, which is in fact consistent with the number of edge modes claimed in the statement of the theorem.

Suppose now that $\omega \leq 0$. The above arguments show that in that case we cannot construct a real mode on the left (because the resulting recursion relation does not admit a normalizable solution). We could perhaps construct an imaginary mode on the left edge, or equivalently, a real mode on the \emph{right} edge. This is true if and only if the spatially inverted system (i.e. where left and right are swapped) admits real zero modes on its left edge. Hence, to prove
\ifx\targetformat\undefined
Theorem 1,
\else
Theorem \ref{thm},
\fi
it is sufficient to prove that if the original system has topological invariant $\omega$, then the inverted system has $\omega_\textrm{inv} = -(\omega + 2c)$. Indeed: if then $\omega_\textrm{inv} > 0$, we can conclude that the original system has $\omega_\textrm{inv} = |\omega+2c|$ imaginary modes on its left edge, whereas if $\omega_\textrm{inv} \leq 0$ it has none. In the latter case we can make the stronger claim that it does not allow for \emph{any} edge modes: if there were a $\gamma$ such that $[\gamma,H] = 0$, we could split it into its real and imaginary components, $\gamma = \frac{\gamma + T\gamma T}{2} + \frac{\gamma - T\gamma T}{2}$, each of which would commute with the Hamiltonian.

We now prove that if $f(z)$ has topological invariant $\omega$, then the inverted chain has a function $f_\textrm{inv}(z)$ with topological invariant $\omega_\textrm{inv} = -\omega - 2c$. First note that spatial inversion comes down to $t_\alpha \leftrightarrow t_{-\alpha}$, and hence $f_\textrm{inv}(z) = f \left( \frac{1}{z} \right)$. If we write $f(z) =  \frac{1}{z^{N_p}} \prod_i (z-z_i)$ and denote the number of zeros of $f(z)$ inside the unit disk as $N_z$ and the \emph{total} number of zeros as $N$, then it is straight-forward to derive that $f \left( \frac{1}{z} \right) \propto \frac{1}{z^{N-N_p}} \prod_i \left(z - \frac{1}{z_i} \right)$. Hence $N^\textrm{inv}_P = N-N_p$. Moreover the total number of zeros is still the same, the number of zeros on the unit circle ($2c$, by definition) has not changed, and the number of zeros of $f \left( \frac{1}{z} \right)$ \emph{outside} the unit disk is the number zeros of $f(z)$ \emph{within} the unit disk, such that $N^\textrm{inv}_Z = N - N_z - 2c$. In conclusion, $\omega_\textrm{inv} = N_z^\textrm{inv} - N_p^\textrm{inv} = N_p - N_z - 2c = -\omega - 2c$. This finalizes the proof.

\textbf{Note.} On first sight the condition that $\omega < -2c$ might seem surprising, however it makes conceptual sense: if $\omega$ satisfies that criterion, then \emph{any} neighboring gapped phase has a winding number that is more negative than $\omega+2c$ (which corresponds to the gapped phase where we move all the zeros on the unit circle inside the unit disk), i.e. then every nearby gapped phase has at least $|\omega+2c|$ edge modes.

\textbf{Example.} Let us treat one example: $H = t_0 H_0 + t_1 H_1 + t_2 H_2$ with homogeneous coordinates $t_0,t_1,t_2 \geq 0$. The associated function is $f(z) = t_2 z^2 + t_1 z + t_0$. It is useful to introduce the inhomogeneous parameters $\tau_0 = \frac{t_0}{t_2}$ and $\tau_1 = \frac{t_1}{t_2}$, such that the two zeros of $f(z)$ are given by $z_{1,2} = - \frac{\tau_1}{2} \pm \sqrt{\left( \frac{\tau_1}{2} \right)^2 - \tau_0} $. Finding the critical points of this model, comes down to characterizing when at least one of the zeros has norm one. For this it is useful to distinguish between when the square root is imaginary or real. In the first case, i.e. $\tau_0 > \left( \frac{\tau_1}{2} \right)^2$, we have $z_{1,2} = - \frac{\tau_1}{2} \pm i \sqrt{\left| \left(\frac{\tau_1}{2} \right)^2 - \tau_0 \right|}$ such that $\left| z_{1,2} \right|^2 = \tau_0$. Hence, one critical line in this model, with \emph{two} zeros on the unit circle, lies at $\tau_0 = 1$ and $\tau_1 < 2$. This is the line $t_0 = t_2$ shown
\ifx\targetformat\undefined
in Fig.~3
\else
in Fig.~\ref{fig:phases}
\fi
of the main text. In the second case, i.e. $\tau_0 \leq \left( \frac{\tau_1}{2} \right)^2$, we have $|z_{1,2}| = \frac{\tau_1}{2} \mp \sqrt{\left( \frac{\tau_1}{2} \right)^2 - \tau_0} $. A straight-forward calculation shows at least one has norm unity if and only if $\tau_1 = \tau_0 + 1$. The system is critical if that condition holds, together with $\tau_0 \leq \left( \frac{\tau_1}{2} \right)^2$, but the former in fact implies the latter. Hence we obtain the critical line $t_1 = t_0 + t_2$ shown
\ifx\targetformat\undefined
in Fig.~3.
\else
in Fig.~\ref{fig:phases}.
\fi

Notice that on the latter line, the zeros of $f(z)$ are $z_1 = -1$ and $z_2 = - \tau_0$. Hence if $\tau_0 <1$, we have a critical point with a Majorana edge mode $\gamma_L = \sqrt{1-\tau_0^2}\sum_{n=1}^\infty (-\tau_0)^{n-1} \gamma_n$ with localization length $\xi = -\frac{1}{\ln \tau_0}$. If $\tau_0 = 0$ we obtain the special case $H = H_1 + H_2$ discussed in the main text, with a perfectly localized edge mode $\gamma_L = \gamma_1$.

\section{Topological invariant in the case of a unit cell} \label{app:unit}

Here we provide more details for the case where the system is not strictly translation invariant, but instead has a repeating unit cell structure. This material is divided into three main sections.

\begin{enumerate}
	\item The first section concerns the \emph{definition} of a Hamiltonian with a unit cell of $N$ sites and its associated complex function $f(z)$ with topological invariant $\omega$. A consistency check is required: if we have a translation invariant system (i.e. $N=1$), the definition of $\omega$ as discussed in the main text can be used, but we are free to describe this system as having a unit cell ($N>1$) and hence choose to use the generalized definition of $\omega$ for systems with a unit cell. We confirm that both options give the same value for $\omega$. Moreover, we show that $\omega$ is a genuine invariant even in the case of a unit cell, i.e. that it cannot change without causing a bulk phase transition. The reasoning is analogous to that of the main text, i.e. the values of $f(z)$ on the unit circle are related to the bulk energy spectrum.
	\item The second section derives the additive property of $\omega$: if one describes two decoupled systems as one system with a larger unit cell, then the $\omega$ thus obtained is simply the sum of the two original topological invariants. This property is shared by the central charge $c$.
	\item The third section concerns the classification of phases in the BDI class. Recall that in the main text we showed that a translation invariant system with central charge $c$ and topological invariant $\omega$ can be tuned to have an associated complex function $f(z) = \pm \left( z^{2c} \pm 1 \right) z^\omega$. We also explained the intuition that if we allow for paths of local Hamiltonians with a unit cell, that we should be able to connect cases which have different signs in the above $f(z)$. Here we confirm this by construction. This completes the proof of Theorem 2 for phases which allow for a translation invariant realization. However, this collection of phases is not closed under stacking: a gapped translation invariant chain stacked on top of a critical translation invariant chain cannot be connected to a single critical translation invariant chain (indeed: the latter never has \emph{any} gapped degrees of freedom). We thus extend the classification to the set of phases generated by stacking. The result is that two systems are in the same phase \emph{if and only if} they have the same $c$ and $\omega$ ---possibly up to a decoupled \emph{trivial} chain (with $c=\omega=0$). The latter is only necessary when one attempts to connect a critical chain \emph{without} gapped degrees of freedom to one \emph{with} them: all the gapped degrees of freedom can then be separated out (i.e. \emph{dumped}) into a decoupled trivial chain.
\end{enumerate}

\subsection{1. Definition of topological invariant}

Suppose that instead of a translation invariant chain, we have a unit cell of $N$ sites. Then our Hamiltonian is of the form $H = \frac{i}{2} \sum_n \bm{\tilde \gamma}_n^T T_\alpha \bm{\gamma}_{n+\alpha}$, where $T_\alpha \in \mathbb R^N\times \mathbb R^N$. The associated meromorphic function we define is
\begin{equation}
f(z) :=  \det F(z) := \det \left( \sum_\alpha T_\alpha z^\alpha \right) .
\end{equation}
We again define the topological invariant as $\omega = N_z - N_p$, where $N_z$ denotes the number of zeros of $f(z)$ strictly within the unit disk, and $N_p$ is the order of the pole at $z=0$.

\subsubsection{$\omega$ is well-defined, i.e. independent of blocking}
A translation invariant system $H = \frac{i}{2} \sum \tilde \gamma_n \; t_\alpha \; \gamma_{n+\alpha}$ has an associated function $g(z) = \sum t_\alpha z^\alpha$. We could also choose to describe this system with an $N$-site unit cell with hopping matrix
\begin{equation}
T_\alpha = \left( \begin{array}{cccc} t_{N\alpha} & t_{N\alpha+1} & \cdots & t_{N\alpha+(N-1)} \\ t_{N\alpha-1} & t_{N\alpha}  & \cdots & t_{N\alpha+(N-2)} \\
\vdots & \vdots & \ddots & \vdots \\
t_{N\alpha - (N-1)} & t_{N\alpha - (N-2)} & \cdots & t_{N\alpha} \end{array} \right),
\end{equation}
with an associated function $f(z) = \det \left( \sum T_\alpha z^\alpha \right)$. We now show that $f(z)$ has the same topological invariant as $g(z)$. To see this, note that we can write
\begin{equation}
f\left( z^N \right) = \det F\left( z^N \right) = \left| \begin{array}{cccc} g_0(z) & g_1(z)/z & \cdots &g_{N-1}(z)/z^{N-1} \\
g_{N-1} (z) \; z & g_0(z) & \cdots &g_{N-2}(z) /z^{N-2} \\
\vdots & \vdots & \ddots & \vdots \\
g_1(z) \; z^{N-1} & g_2(z) \; z^{N-2} & \cdots & g_0(z) \end{array} \right|
= 
\left| \begin{array}{cccc}
g_0(z) & g_1(z) & \cdots &g_{N-1}(z) \\
g_{N-1} (z) & g_0(z) & \cdots &g_{N-2}(z) \\
\vdots & \vdots & \ddots & \vdots \\
g_1(z) & g_2(z) & \cdots & g_0(z) \end{array} \right|,
\end{equation}
where $g_i(z) = \sum_n t_{n N + i} z^{nN+i}$, giving a decomposition $g(z) = \sum_i g_i(z)$ such that $g_i\left( \zeta z \right) = \zeta^i \; g_i(z)$ where $\zeta = e^{2\pi i/N}$. The above means that $f\left( z^N \right)$ is the determinant of the circulant matrix associated to the list of numbers $\left(g_0(z), \cdots, g_{N-1}(z) \right)$, whose determinant is known to be given by $\prod_{n=0}^{N-1} \sum_i \zeta^{ni} \; g_i(z)$. Hence $f\left( z^N \right) = \prod_{n=0}^{N-1} \sum_i g_i\left(\zeta^n z\right) = \prod_{n=0}^{N-1} g\left(\zeta^n z \right)$. Note that the topological invariant of $f\left(z^N\right)$ is simply $N$ times that of $f(z)$, hence we conclude $f(z)$ and $g(z)$ have the same topological invariant. This means $\omega$ does not depend on how we \emph{choose} to describe our system.

As an aside, it is worth noting that $f(z)$ and $g(z)$ do \emph{not} coincide. For example, $H_0 + H_2$ has the associated function $g(z) = 1+z^2 = (z+i)(z-i)$, whereas the above shows that $f(z^2) = g(z) g(-z) = (1+z^2)^2$, such that $f(z) = (1+z)^2$. Observe that $f(z)$ has a zero at $z=-1$ with multiplicity \emph{two}. Unlike in the translation invariant case, this no longer implies that the dynamical critical exponent $z_\textrm{dyn} = 2$. This illustrates that if the system has a unit cell, we can no longer use the associated function to distinguish between, for example, a CFT with $c=1$ or a single quadratic gapless point. However, it remains true that \emph{if the bulk is a CFT}, then the central charge is given by half the number of zeros on the unit circle (counting multiplicities). This is a consequence of the relation between $f(e^{ik})$ and the energy spectrum, which we prove now.

\subsubsection{$\omega$ cannot change without a phase transition}

The values of $f(z)$ on the unit circle carry the same relevance as in the translation invariant case. In particular, a zero crossing the unit circle would correspond to changing the physics in the bulk, since $\left|f\left( e^{ik} \right)\right| = \prod_{n=1}^N \left|\varepsilon_k^{(n)}\right|$, where $\varepsilon_k^{(n=1,\cdots,N)}$ represent the $N$ bands. To see this, consider the Hamiltonian with an $N$-site unit cell:
\begin{equation}
H = \frac{i}{2} \sum_\alpha \sum_n \left( \tilde \gamma_{n,1} \quad \tilde \gamma_{n,2}\quad \cdots \quad \tilde \gamma_{n,N} \right) \; T_\alpha \;  \left(\begin{array}{c} \gamma_{n+\alpha,1} \\ \gamma_{n+\alpha,2} \\
\vdots \\ \gamma_{n+\alpha,N} \end{array}\right).
\end{equation}
If we define $c_{k,\lambda} = \sqrt{\frac{N}{L}} \sum_n e^{-ikn} c_{n,\lambda}$ (where $\lambda$ is the index within the unit cell and $L$ is the \emph{total} number of sites), then a straight-forward computation shows that
\begin{equation}
H = - \frac{1}{2}\sum_k \left( c_{k,1}^\dagger \quad \cdots \quad c_{k,N}^\dagger \quad c_{-k,1} \quad \cdots \quad c_{-k,N} \right) H_k \left( \begin{array}{l} c_{k,1} \\ \vdots \\ c_{k,N} \\ c_{-k,1}^\dagger \\ \vdots \\ c_{-k,N}^\dagger \end{array} \right), \; \; \; \; H_k = \left( \begin{array}{cc}
F_H \left( z \right) & F_{AH} \left( z \right) \\
- F_{AH} \left(z \right) & - F_H \left( z \right)
\end{array} \right) \Bigg|_{z = e^{ik}},
\end{equation}
where $F_H(z)$ and $F_{AH}(z)$ denote the Hermitian and anti-Hermitian part of $F(z) = \sum_\alpha T_\alpha z^\alpha$, i.e. $\frac{F(z) \pm F(z)^\dagger}{2}$. The determinant of $H_k$ is most easily obtained by conjugating with $U = \frac{1}{\sqrt{2}} (\sigma_x+\sigma_z) \otimes \mathbb I_N$:
\begin{equation}
U H_k U^\dagger = \left( \begin{array}{cc} 0 & F(z)^\dagger \\
F(z) & 0 \end{array} \right) \Bigg|_{z = e^{ik}}
\end{equation}
such that $\det H_k = (-1)^N \left|f\left( e^{ik} \right)\right|^2$. This finishes the proof since $\det H_k = (-1)^N \prod_{n=1}^N \left( \varepsilon_k^{(n)} \right)^2$.

Hence even in the presence of a unit cell, we can associate a meromorphic function to the Hamiltonian. Moreover, this gives rise to a well-defined topological invariant (independent of whether the system is gapped or gapless) that can only change if the bulk undergoes a phase transition.

\subsection{2. Additivity of $c$ and $\omega$ under stacking}

Take two chains, characterized by $g(z) = \det G(z) = \det \left( \sum T^{(1)}_\alpha z^\alpha\right)$ and $h(z) = \det H(z) = \det \left( \sum T^{(2)}_\alpha z^\alpha\right)$ respectively. One can stack these two on top of one another, which is equivalent to having a chain with
\begin{equation}
T_\alpha = \left( \begin{array}{cc} T^{(1)}_{\alpha} & 0 \\0 & T^{(2)}_{\alpha} \end{array} \right) \qquad \textrm{such that }
f(z) = \left|\begin{array}{cc} G(z) & 0 \\
0 & H(z) \end{array} \right| = g(z) h(z) \; .
\end{equation}
Hence the topological invariant of $f(z)$ is the sum of those of $g(z)$ and $h(z)$. This establishes the additive property of the topological invariant under stacking ---similar to the well-known property of the central charge of a CFT. Note that the central charge can only increase under addition, such that the \emph{labeling} by $\mathbb N \times \mathbb Z$, with $c \in \frac{1}{2} \mathbb N$ and $\omega \in \mathbb Z$, is in fact a \emph{semigroup}.

In the main text we saw that $c$ and $\omega$ \emph{uniquely} label all the phases which allow for at least one translation invariant representative (if translation symmetry is not enforced). Stacking such phases can in principle generate new phases, and hence it is a priori not clear that $c$ and $\omega$ are sufficient to label them, i.e. that stacks with the same $c$ and $\omega$ can be connected. In the next section we show that this is in fact true. This means that the set of phases can be \emph{identified} with the semigroup $\mathbb N \times \mathbb Z$, where the operation of stacking corresponds to addition. The special case where $c = 0$ corresponds to the classification of gapped phases in the BDI class which is known to be classified by the \emph{group} $\mathbb Z$, which in this context can be identified with $\{0\} \times \mathbb Z \subset \mathbb N \times \mathbb Z$.

\subsection{3. Classification}

\subsubsection{Effect of unit cell on classification of translation invariant chains}

\begin{figure}[h]
	\includegraphics[scale=.8]{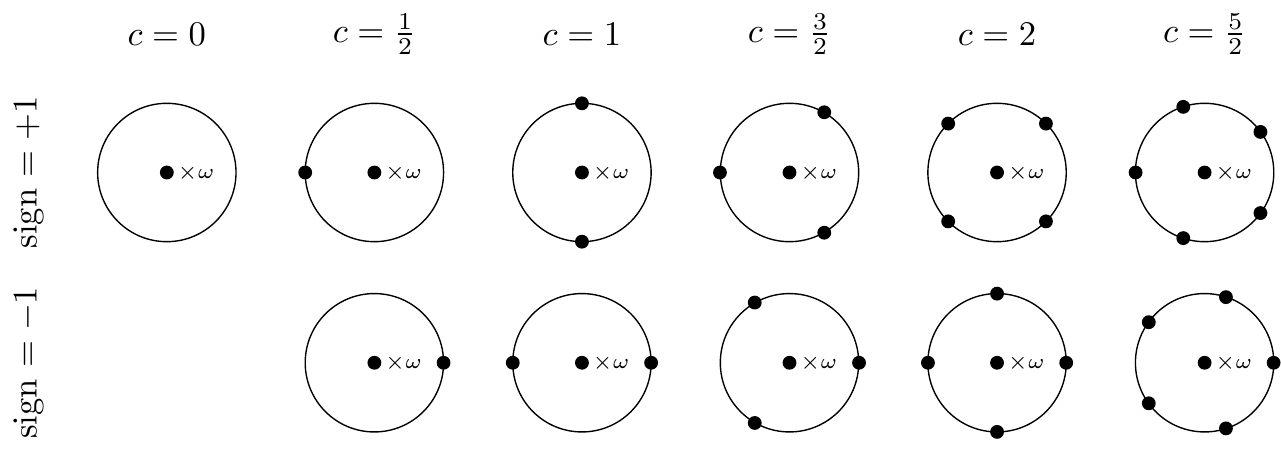}
	\caption{Any translation invariant Hamiltonian in the BDI class, $H = \sum t_\alpha H_\alpha$ with $f(z) = \sum t_\alpha z^\alpha$, which has central charge $c \in \frac{1}{2} \mathbb N$ and topological invariant $\omega \in \mathbb Z$, can be brought to the canonical form $H = \pm \left( H_{2c+\omega} \pm H_{\omega} \right)$ with $f(z) = \pm \left(z^{2c} \pm 1 \right) z^\omega$ without causing a phase transition (as explained in the main text). The pattern of zeros and poles of this canonical form is illustrated above, where `sign' denotes the choice of the relative sign in $z^{2c} \pm 1$. Including the further choice of the overall sign, one obtains a $\mathbb Z_2 \times \mathbb Z$ classification for the gapped phases ($c=0$), and a $\mathbb Z_2 \times \mathbb Z_2 \times \mathbb Z$ classification for the critical phases which are described by a non-trivial CFT in the bulk. If we allow for paths which break strict translation symmetry by introducing a unit cell, we can always bring it to the form $H = H_{2c+\omega} + H_{\omega}$: the central charge and the topological invariant fully characterize each phase (as long as the dynamical critical exponent $z_\textrm{dyn} = 1$, i.e. the zeros on the unit circle are non-degenerate). \label{fig:classification}}
\end{figure}

In the main text we explained how any translation invariant model with central charge $c$ and topological invariant $\omega$ can be tuned to the canonical form $f(z) = \pm \left(z^{2c} \pm 1 \right) z^\omega$ whilst at all times preserving translation invariance. From the above we know that even if we allow for paths with an arbitrary unit cell, $\omega$ \emph{cannot} change without causing a phase transition (whereas $c$ is not allowed to change by virtue of our very \emph{definition} of a phase). However, we now show that by allowing for such a unit cell, we \emph{can} connect models with such different signs. The situation is summarized in Fig.~\ref{fig:classification}.

Regarding the first sign, one can note that defining $U(\alpha) = \exp\left(\frac{\alpha}{2} \sum_n \tilde \gamma_{2n-1} \tilde \gamma_{2n}\right) = \prod_n \left( \cos\left( \frac{\alpha}{2} \right) + \tilde \gamma_{2n-1} \tilde \gamma_{2n} \sin\left( \frac{\alpha}{2} \right) \right)$ does the trick: $U(\pi) H_\alpha U(\pi)^\dagger = -H_\alpha$ whereas $[U(\alpha),T] = 0$. Although $U(\pi)$ maps strictly translation invariant models to strictly translation invariant models, for intermediate $0<\alpha<\pi$, the transformed Hamiltonian will have a two-site unit cell. 

For the second sign, we need to find a transformation which maps $H_{\omega} \to - H_{\omega} $ and $H_{2c+\omega} \to H_{2c+\omega}$. Note that at a discrete level, this is accomplished by applying (for a chain with periodic boundary conditions)
\begin{equation}
V =  \quad 
\cdots \left(\tilde \gamma_1 \cdots \tilde \gamma_{2c} \right) \left(\gamma_{2c+1+\omega} \cdots \gamma_{4c+\omega} \right)
\left(\tilde \gamma_{4c+1} \cdots \tilde \gamma_{6c} \right)
\left(\gamma_{6c+1+\omega} \cdots \gamma_{8c+\omega} \right) \cdots
\end{equation}
Indeed: $V H_\omega V^\dagger = - H_\omega$ and $V H_{2c+\omega} V^\dagger = H_{2c+\omega}$. We now need to show that this can be implemented gradually. To this purpose, define the local generator
\begin{equation}
A = \sum_{n \in \mathbb Z} \sum_{m=1}^{2c} \tilde \gamma_{6nc+m} \tilde \gamma_{6nc+4c+m} + \gamma_{6nc+2c+\omega+m} \gamma_{6(n+1)c+\omega+m} \; .
\end{equation}
Firstly note that $[A,T] = 0$. Secondly, all the terms in $A$ commute, such that
\begin{align}
V(\alpha) &:= \exp \left( \frac{\alpha}{2} A \right) \\
&= \prod_{n\in\mathbb Z}\prod_{m=1}^{2c}  \exp \left( \frac{\alpha}{2} \tilde \gamma_{6nc+m} \tilde \gamma_{6nc+4c+m} \right) \; \exp \left( \frac{\alpha}{2} \gamma_{6nc+2c+\omega+m} \gamma_{6(n+1)c+\omega+m} \right) \\
&= \prod_{n\in\mathbb Z}\prod_{m=1}^{2c} \left( \cos\left( \frac{\alpha}{2} \right) +  \tilde \gamma_{6nc+m} \tilde \gamma_{6nc+4c+m} \sin\left( \frac{\alpha}{2} \right) \right) \; \left( \cos\left( \frac{\alpha}{2} \right) +  \gamma_{6nc+2c+\omega+m} \gamma_{6(n+1)c+\omega+m} \sin\left( \frac{\alpha}{2} \right) \right).
\end{align}
Hence this allows to smoothly apply $V$, since 
\begin{equation}
V(\pi) = \prod_{n\in\mathbb Z}\prod_{m=1}^{2c} \tilde \gamma_{6nc+m} \tilde \gamma_{6nc+4c+m} \gamma_{6nc+2c+\omega+m} \gamma_{6(n+1)c+\omega+m} = V \; .
\end{equation}

\subsubsection{Extending the classification to stacks: a semigroup structure}

We consider all phases in the BDI class which are described in the bulk by a CFT and which can be obtained by deforming a translation invariant Hamiltonian $H = \sum t_\alpha H_\alpha$ ---or stackings thereof--- with an arbitrary unit cell. We now prove that these are classified by the semigroup $\mathbb N \times \mathbb Z$, being uniquely labeled by the central charge $c \in \frac{1}{2} \mathbb N$ and the topological invariant $\omega \in \mathbb Z$. More precisely, given such a system, then after adding a \emph{single} trivial decoupled chain $H_0$ \emph{if need be}, we can always smoothly connect it to a chain described by $H_\omega$ stacked on top of $2c$ copies of the standard critical Majorana chain $H_0 + H_1$.

Note that given the previous section, it is sufficient to prove the above statement for a system consisting of $n$ decoupled chains, each in the canonical form $H_i = H_{2c_i+\omega_i} + H_{\omega_i}$ with central charge $c_i$ and topological invariant $\omega_i$ such that $c = \sum_{i=1}^n c_i$ and $\omega = \sum_{i=1}^n \omega_i$. We prove this by induction.

Consider the case $n=1$. If $c = 0$, then the system is already described by $H_\omega$ and we are done. If $c \neq 0$, then after stacking with a trivial chain described by $H_0$, we can bring it to two decoupled chains $H_\omega$ and $H_0 + H_{2c}$ by virtue of Corollary~\ref{cor} (an auxilliary result proved below). The latter can naturally be seen as consisting of $2c$ decoupled copies of $H_0 + H_1$.

As for the induction step: suppose the decomposition holds for $n$, then we prove it for $n+1$. Applying the result on the first $n$ chains gives us $H_{\omega_1+\cdots+\omega_n}$ and $2(c_1+\cdots+c_n)$ copies of $H_0 + H_1$. We can now use Corollary~\ref{cor} to transfer the topological invariant of the $(n+1)$th chain into $H_{\omega_1+\cdots+\omega_n}$, leaving the $(n+1)$th chain to be described by $H_0 + H_{2c_{n+1}}$. If $c_{n+1} \neq 0$, the latter consists of $2c_{n+1}$ copies $H_0 + H_1$ such that we are done. If $c_{n+1} = 0$, then we end up with \emph{two} decoupled gapped chains (in addition to the critical chains), described by $H_\omega$ and $H_0$. However, such a stack can locally and smoothly be connected into a single chain $H_\omega$. One way of seeing this, is by using Corollary~\ref{cor} to connect this to two copies of $H_{\frac{\omega}{2}}$ if $\omega$ is even, or to a stack of $H_{\frac{\omega-1}{2}}$ and $H_{\frac{\omega+1}{2}}$ if $\omega$ is odd. In the former case, simply interlacing the two chains gives a chain described by $H_\omega$, whereas in the latter case the same works if one also swaps the corresponding real Majoranas of both chains. This swapping can be done locally, smoothly and in a $T$-preserving way by virtue of Lemma~\ref{lemma}, which is proven below. This completes the proof.

\begin{lemma} \label{lemma}
	A permutation of a finite set of Majorana modes of the same type (i.e. real or imaginary) can be implemented in a smooth and $T$-preserving fashion, on the condition that the same permutation cycle is applied on two disjoint systems.
	
	Note that a permutation of a finite number of Majoranas is automatically local. If one is interested in non-overlapping permutations cycles (whose cycle length does not depend on system size) in an arbitrarily large system, then by applying this result in parallel, one can achieve this in a local, smooth and $T$-preserving fashion.
\end{lemma}
Since any permutation can be decomposed into a series of transpositions, it is sufficient to prove this lemma for the special case of a swap $\gamma_1 \leftrightarrow \gamma_2$. The condition that the same permutation cycle is applied on a disjoint system, means we also want to swap $\gamma_3 \leftrightarrow \gamma_4$. To this end, let us define
\begin{align}
U(\alpha) &= \exp \left( \frac{\alpha}{2} \gamma_2\gamma_4 \right) \exp \left( \frac{\alpha}{4}\left[  \gamma_1\gamma_2 + \gamma_3\gamma_4 \right]\right) \\
&= \left[\cos\left( \frac{\alpha}{2} \right) + \gamma_2\gamma_4 \sin\left( \frac{\alpha}{2} \right) \right]\left[\cos\left( \frac{\alpha}{4} \right) + \gamma_1\gamma_2 \sin\left( \frac{\alpha}{4} \right) \right] \left[\cos\left( \frac{\alpha}{4} \right) + \gamma_3\gamma_4 \sin\left( \frac{\alpha}{4} \right) \right].
\end{align}
It is clear that for $\alpha \in \mathbb R$ we have $[U(\alpha),T]=0$, and the Hermiticity of the Majorana modes shows that $U(\alpha)$ is unitary. It is straightforward to show that $U(\pi) = \frac{1}{2} (\gamma_2-\gamma_1)(\gamma_4-\gamma_3)$. This means that $U(\pi) \gamma_1 U(\pi)^\dagger = \gamma_2$ and $U(\pi) \gamma_2 U(\pi)^\dagger = \gamma_1$ (and analogously for $\gamma_3$ and $\gamma_4$). Hence, $U(\alpha)$ indeed smoothly implements the two swaps for $\alpha \in [0,\pi]$.

Note that we needed to simultaneously do the swap on $\gamma_3,\gamma_4$ as well, otherwise we can only define $V(\alpha) = \exp \left( \frac{\alpha}{4} \gamma_1\gamma_2 \right)$, which indeed achieves $V(\pi) \gamma_2 V(\pi)^\dagger= \gamma_1$ but also $V(\pi) \gamma_1 V(\pi)^\dagger= -\gamma_2$.

We now show that this lemma implies a very useful corollary.

\begin{corollary} \label{cor}
	Given two decoupled translation invariant chains in the BDI class with topological invariants $\omega_1$ and $\omega_2$ respectively, there is a local, smooth and $T$-preserving way to shift $\omega_1 \to \omega_1 + \zeta$ and $\omega_2 \to \omega_2 - \zeta$ for arbitrary $\zeta \in \mathbb Z$. 
	
	Colloquially, we say that we can transfer the topological invariant between the two chains whilst staying within the same phase.
\end{corollary}

Note that for a \emph{single} chain we can arbitrarily modify its topological invariant by shifting each real mode $\gamma_n \to \gamma_{n+\zeta}$, which has the effect $t_\alpha \to t_{\alpha + \zeta}$ such that $f(z) \to z^\zeta f(z)$ and hence $\omega \to \omega +\zeta$. However, this discrete mapping cannot be implemented in local, smooth and $T$-preserving way. Indeed, we know we cannot change $\omega$ without causing a phase transition.

\begin{figure}[h]
	\includegraphics[scale=1]{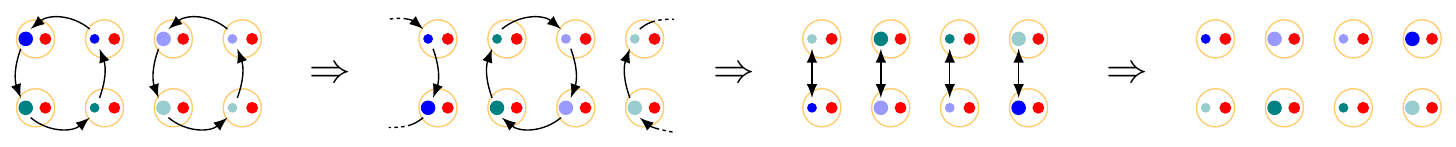}
	\caption{A visual proof of the statement that by \emph{local} permutations of Majoranas of the \emph{same} type (blue and green denote \emph{real} Majorana modes, red denote \emph{imaginary} modes) one can map $\gamma_n \to \gamma_{n-1}$ in the \emph{top} layer and $\gamma_n \to \gamma_{n+1}$ in the \emph{bottom} layer (and one can reverse the shift by changing the direction of the arrows in the figure). This implies that if the two chains were decoupled, then under this transformation $H^\textrm{(top)} = \sum t^\textrm{(top)}_\alpha H_\alpha^\textrm{(top)} \to \sum t^\textrm{(top)}_\alpha H_{\alpha-1}^\textrm{(top)}$ (and in the opposite direction for $H^\textrm{(bottom)}$). In other words: this allows to transfer the topological invariant from one chain to the other (whilst keeping the total $\omega$ unchanged). As a consequence of Lemma~\ref{lemma}, this type of transformation can be done smoothly in a local and $T$-preserving way. Hence this allows, for example, to continuously connect a stack of $H_0$ ($\omega=0$, $c=0$) and $H_1+ H_2$ ($\omega=1$, $c=\frac{1}{2}$) to a stack of $H_1$ ($\omega=1$, $c=0$) and $H_0+H_1$ ($\omega=0$, $c=\frac{1}{2}$).\label{fig:shift}}
\end{figure}

However, given a stack of two chains, it turns out we can implement the above shift map (in opposite directions on each chain, such that the total topological invariant is unaffected) via a series of \emph{local} permutations. The corollary then follows by applying Lemma~\ref{lemma}. The necessary permutations are shown in Fig.~\ref{fig:shift}.

\end{document}